\documentclass[nofootinbib,aps,11pt]{revtex4-1}
\usepackage{graphicx}
\usepackage{hyperref}
\usepackage{cancel}
\usepackage{amssymb}
\usepackage{textcomp}
\usepackage{amsmath}
\usepackage{bm}
\usepackage{times}
\usepackage{epsfig}
\usepackage{color}
\usepackage{multirow}
\usepackage{subfigure}

\begin{document}
\title{\LARGE Probing Dirac Neutrino Properties with Dilepton Signature}
\bigskip
\author{Wan-Lun Xu$^1$}
\author{Zhi-Long Han$^1$}
\email{sps\_hanzl@ujn.edu.cn}
\author{Yi Jin$^{1,2}$}
\author{Honglei Li$^1$}
\author{Zongyang Lu$^1$}
\author{Zhao-Xia Meng$^1$}
\email{sps\_mengzx@ujn.edu.cn}
\affiliation{
	$^1$School of Physics and Technology, University of Jinan, Jinan, Shandong 250022, China
	\\
	$^2$Guangxi Key Laboratory of Nuclear Physics and Nuclear Technology, Guangxi Normal University, Guilin, Guangxi 541004, China}
\date{\today}

\begin{abstract}
	The neutrinophilic two Higgs doublet model is one of the simplest models to explain the origin of tiny Dirac neutrino masses. This model introduces a new Higgs doublet with eV scale VEV to naturally generate the tiny neutrino masses. Depending on the same Yukawa coupling, the neutrino oscillation patterns can be probed with the dilepton signature from the decay of charged scalar $H^\pm$. For example, the normal hierarchy predicts BR$(H^+\to e^+\nu)\ll$ BR$(H^+\to \mu^+\nu)\approx$ BR$(H^+\to \tau^+\nu)\simeq0.5$ when the lightest neutrino mass is below 0.01 eV, while the inverted hierarchy predicts BR$(H^+\to e^+\nu)/2\simeq$ BR$(H^+\to \mu^+\nu)\simeq$ BR$(H^+\to \tau^+\nu)\simeq0.25$. By precise measurement of  BR$(H^+\to \ell^+\nu)$, we are hopefully to probe the lightest neutrino mass and the atmospheric mixing angle $\theta_{23}$.  Through the detailed simulation of the dilepton signature and corresponding backgrounds, we find that the 3 TeV CLIC could discover $M_{H^+}\lesssim1220$ GeV for NH and $M_{H^+}\lesssim1280$ GeV for IH. Meanwhile, the future 100 TeV FCC-hh collider could probe $M_{H^+}\lesssim1810$ GeV for NH and $M_{H^+}\lesssim2060$ GeV for IH.
\end{abstract}

\maketitle

\section{Introduction}

The neutrino oscillation experiments have confirmed that neutrinos have non-zero masses \cite{Super-Kamiokande:1998kpq,SNO:2002tuh}, which is concrete evidence for new physics beyond the standard model (SM). With great effort made by the neutrino oscillation experiments in the last two decades, the picture of three neutrino oscillations has become more and more precise \cite{Esteban:2020cvm}. Currently, one major issue of the neutrino oscillation is the neutrino mass hierarchy, where the sign of $\Delta m_{31}^2=m_{\nu_3}^2-m_{\nu_1}^2$ is undetermined. The combined analysis of Super-Kamiokande \cite{Super-Kamiokande:2017yvm}, T2K \cite{T2K:2021xwb} and NO$\nu$A \cite{NOvA:2021nfi} experiments shows possible hints for normal ordering at present\cite{Esteban:2020cvm}. This hierarchy problem may be solved by future oscillation experiments such as DUNE \cite{DUNE:2020ypp} and JUNO \cite{JUNO:2015zny}. The other two quantities that need to be measured are the octant of the atmospheric mixing angle $\theta_{23}$ and the CP violation phase $\delta_\text{CP}$ \cite{deSalas:2020pgw}.

Because the oscillation experiments are only sensitive to the squared mass differences of neutrinos, they can not tell the absolute neutrino mass scale either. Meanwhile, the massive neutrinos could lead to sizable impacts on the cosmic microwave background and the large scale structure \cite{Dvorkin:2019jgs}, which has set an upper bound on the sum of light neutrino masses as  $\sum m_\nu<0.12$~eV \cite{Planck:2018vyg}. Therefore, neutrino masses are below  the eV scale. Although the cosmological limit is model-dependent \cite{DiValentino:2021hoh}, it is more stringent than the direct kinematic measurement \cite{KATRIN:2019yun}.

Besides the above unknowns, there are still many questions about neutrinos to be resolved. The crucial one is the Dirac or Majorana nature of the neutrino. Provided the neutrinos are Majorana type, then the lepton number is violated, which can be further confirmed by neutrinoless double beta decay \cite{Dolinski:2019nrj}. Since no positive lepton number violation signature has been observed until now, the Dirac type neutrino is also possible. Then we have to consider the origin of the sub-eV scale Dirac neutrino mass. Of course, the Yukawa interaction with SM Higgs doublet as $y_\nu \bar{L}\tilde{\Phi}_1\nu_R$ can generate the tiny neutrino mass with feeble coupling $y_\nu\sim10^{-12}$, which seems unnatural small compared with other SM fermions. The natural generation of Dirac neutrino mass is also studied extensively \cite{Gu:2006dc,Farzan:2012sa,Ma:2016mwh,Wang:2016lve,Yao:2017vtm,Borah:2018gjk,Saad:2019bqf,Guo:2020qin,Calle:2021tez,Chen:2022bjb}. In this paper, we consider the neutrinophilic two Higgs doublet model ($\nu$2HDM) \cite{Davidson:2009ha,Davidson:2010sf},  where a new Higgs doublet $\Phi_2$ with a tiny vacuum expectation value (VEV) generates the Dirac neutrino mass.

In canonical seesaw models for Majorana neutrino, there is a strong connection between the branching ratio of heavy new particles and the neutrino oscillation parameters \cite{Akeroyd:2007zv,FileviezPerez:2009hdc}. Once these new particles are discovered at colliders, the measurement of corresponding branching ratios provides an alternative pathway to probe neutrino mixing parameters \cite{Mandal:2022ysp,Mandal:2022zmy}. In the $\nu$2HDM, the Dirac neutrino mass matrix is proportional to the product of the neutrinophilic Yukawa coupling $y_\nu$ and the VEV $v_2$. Since the decays of the new charged scalar $H^\pm$ are also determined by the coupling $y_\nu$, the branching ratios of $H^\pm$ are expected to depend on the neutrino oscillation parameters \cite{Davidson:2009ha,Davidson:2010sf}. In this paper, we study the capability of future colliders such as the CLIC \cite{CLIC:2018fvx} and FCC-hh \cite{Arkani-Hamed:2015vfh} to probe the Dirac neutrino properties with dilepton signature. This signature is hopeful to unravel the neutrino mass hierarchy, the atmospheric octant, and the neutrino mass scale by precise measurements of BR$(H^\pm\to \ell^\pm \nu)$.

The paper is organized as follows. In Section \ref{SEC:MD}, we briefly review the $\nu$2HDM for Dirac neutrino mass and relevant constraints. The branching ratio of charged scalar $H^\pm$ and its relation with the neutrino oscillation parameters are discussed in Section \ref{SEC:BR}. The dilepton signature at the 3 TeV CLIC and 100 TeV FCC-hh are studied in Section \ref{SEC:CLIC} and Section \ref{SEC:hh} respectively. The conclusion is given in Section \ref{SEC:CL}.

\section{The $\nu$2HDM Model}\label{SEC:MD}
 Besides the SM Higgs doublet $\Phi_1$, this model further introduces one Higgs doublet $\Phi_2$ and three right-handed neutrinos $\nu_{R}$. To forbid the direct Yukawa interaction between SM Higgs $\Phi_1$ and right-handed neutrino $\nu_R$, a global $U(1)$ symmetry is also imposed. Under this $U(1)$ symmetry, both the new Higgs doublet $\Phi_2$ and right-handed neutrino $\nu_R$ carry charge $+1$, while all SM fields remain uncharged. The Majorana mass term of right-handed neutrino $M_R \overline{\nu^c_R}\nu_R$ is also forbidden by the $U(1)$ symmetry. The two Higgs doublets can be expressed in the form of 
 \begin{align}
 	\Phi_i=\left(
 	\begin{array}{c}
 		\phi_i^+\\
 		(v_i+\phi_i^{0,r}+i \phi_i^{0,i})/\sqrt{2}
 	\end{array}\right),
 \end{align}
with $v_i$ the corresponding VEV. The scalar potential under the global $U(1)$ symmetry is
 \begin{eqnarray}
 	V&=&m_{11}^2\Phi_1^\dag\Phi_1+m_{22}^2\Phi_2^\dag\Phi_2-\left[m_{12}^2\Phi_1^\dag\Phi_2+\text{h.c.}\right]+\frac{\lambda_1}{2}\left(\Phi_1^\dag\Phi_1\right)^2 
 	\\ \nonumber
 	&&+\frac{\lambda_2}{2}\left(\Phi_2^\dag\Phi_2\right)^2 +\frac{\lambda_3}{2}\left(\Phi_1^\dag\Phi_1\right)\left(\Phi_2^\dag\Phi_2\right) +\frac{\lambda_4}{2}\left(\Phi_1^\dag\Phi_2\right)\left(\Phi_2^\dag\Phi_1\right),
 \end{eqnarray}
where the soft term $m_{12}^2\Phi_1^\dag\Phi_2$ breaks the global $U(1)$ symmetry explicitly. Because a vanishing $m_{12}^2$ would restore the global $U(1)$ symmetry, it is naturally small \cite{tHooft:1979rat}. The electroweak symmetry breaking is triggered spontaneously by $m_{11}^2<0$, while $m_{22}^2+(\lambda_3+\lambda_4)v_1^2/2>0$ is required to prevent the spontaneous breaking of $U(1)$ symmetry. The soft term $m_{12}^2\Phi_1^\dag\Phi_2$ induces a tiny VEV for $\Phi_2$ as
\begin{equation}
	v_2 = \frac{m_{12}^2 v_1}{m_{22}^2+(\lambda_3+\lambda_4)v_1^2/2}.
\end{equation}
Then $v_2\sim$ eV can be obtained with $m_{12}^2\sim\text{MeV}^2$. The relation $v_2\ll v_1$ is stable under radiative corrections, because the corrections are only logarithmically sensitive to the cutoff \cite{Morozumi:2011zu,Haba:2011fn}. To ensure the stability of the potential at large field values, one needs
\begin{equation}
	\lambda_{1,2}>0,\lambda_3+\sqrt{\lambda_1\lambda_2}>0,\lambda_3+\lambda_4+\sqrt{\lambda_1\lambda_2}>0.
\end{equation}

The physical Higgs bosons are given by \cite{Guo:2017ybk}
\begin{eqnarray}
	H^+=\phi^+_2\cos\beta-\phi^+_1\sin\beta&,~& A=\phi^{0,i}_2\cos\beta-\phi^{0,i}_1\sin\beta, \\
	H=\phi^{0,r}_2\cos\alpha-\phi^{0,r}_1\sin\alpha&,~& h=\phi^{0,r}_1\cos\alpha+\phi^{0,r}_2\sin\alpha,
\end{eqnarray}
with the mixing angles $\tan\beta=v_2/v_1$ and $\tan2\alpha\simeq 2 v_2/v_1$. Since the mixing angles are heavily suppressed by the tiny VEV $v_2$, the new Higgs bosons are almost the neutrinophilic doublet. Masses of the scalars are
\begin{eqnarray}
	M_{H^+}^2\simeq m_{22}^2\!+\frac{1}{2}\lambda_3v_1^2,~M_A^2\simeq M_H^2\simeq M_{H^+}^2\!+\frac{1}{2}\lambda_4v_1^2,~ M_h^2 \simeq  \lambda_1 v_1^2,
\end{eqnarray}
where terms of order $m_{12}^2$ and $v_2^2$ are neglected. The charged scalar can be pair produced at colliders and leads to the dilepton signature $\ell^+\ell^-+P_T^{miss}$ \cite{Davidson:2009ha,Davidson:2010sf}. The direct searches for slepton at LHC in the same dilepton signature  have excluded $M_{H^+}\lesssim 700$~GeV \cite{ATLAS:2019lff,CMS:2020bfa}.

Under the global $U(1)$ symmetry, the allowed new Yukawa interaction is
\begin{equation}
	\mathcal{L}_Y=-y_\nu  \bar{L}\tilde{\Phi}_2\nu_R + \text{h.c.},
\end{equation}
which induces the tiny neutrino mass as $m_\nu=y_\nu v_2/\sqrt{2}$. The Yukawa couplings of the new scalars in the neutrino mass eigenstate are
\begin{equation}
	\mathcal{L}_Y=-\frac{m_{\nu_i}}{v_2} H \bar{\nu}_i \nu_i + \frac{m_{\nu_i}}{v_2} A \bar{\nu}_i \gamma_5\nu_i- \frac{\sqrt{2} m_{\nu_i}}{v_2}\left(U_{\ell i}^{*} H^+ \bar{\nu}_i \ell_L + \text{h.c.}\right),
\end{equation}
where $U$ is  the Pontecorvo-Maki-Nakagawa-Sakata (PMNS) matrix. The PMNS matrix can be written explicitly as \cite{Gluza:2001de}
\begin{align}
	U\! =\! \left(
	\begin{array}{ccc}
		c_{12} c_{13} & s_{12} c_{13} & s_{13} e^{i\delta}\\
		-s_{12}c_{23}-c_{12}s_{23}s_{13}e^{-i\delta} & c_{12}c_{23}-s_{12}s_{23}s_{13} e^{-i\delta} & s_{23}c_{13}\\
		s_{12}s_{23}-c_{12}c_{23}s_{13}e^{-i\delta} & -c_{12}s_{23}-s_{12}c_{23}s_{13}e^{-i\delta} & c_{23}c_{13}
	\end{array}
	\right)
\end{align}
where $c_{ij}\equiv \cos\theta_{ij}$ and $s_{ij}\equiv\sin\theta_{ij}$ for short, and $\delta$ is the CP violation phase. In this paper, we consider the latest results for three neutrino oscillations \cite{Esteban:2020cvm}.

The relatively large Yukawa coupling of charged scalar could induce the lepton flavor violation process \cite{Bertuzzo:2015ada}. Currently, the most stringent constraint comes from the MEG collaboration, which is BR$(\mu\to e\gamma)<4.2\times10^{-13}$ at 90\% CL \cite{MEG:2016leq}. Its branching ratio is calculated as \cite{Fukuyama:2008sz}
\begin{equation}
	\text{BR}(\mu\to e\gamma)=\frac{\alpha}{24\pi} \left(\frac{v_1}{v_2}\right)^4 \frac{|m_{\nu_j}^2U_{ej}U_{\mu j}|^2}{M_{H^+}^4}.
\end{equation}
The experimental upper limit sets a lower bound on $v_2$ as
\begin{equation}
	v_2\gtrsim 15 \text{eV} \left(\frac{m_\nu}{0.05\text{eV}}\right)\left(\frac{100\text{GeV}}{M_{H^+}}\right).
\end{equation}
Under the direct constraints from the collider, $v_2\gtrsim \mathcal{O}(\text{eV})$ is allowed for $M_{H^+}\sim \mathcal{O}(\text{TeV})$.

\section{Branching Ratio of $H^\pm$}\label{SEC:BR}

Because there are strong correlations between the neutrino oscillation parameters and the branching ratio of charged scalar $H^\pm$, we perform a detailed analysis in this section. For eV scale $v_2$, the partial decay width as $H^+\to tb$ is heavily suppressed by the mixing angle $\tan\beta=v_2/v_1\sim10^{-11}$, thus it can be neglected.  Meanwhile, the cascade decay $H^+\to W^+H/W^+A$ can be kinematically forbidden provided $M_{H,A}\ge  M_{H^+}$, which is realized by requiring $\lambda_4>0$. In this way, $H^+\to\ell^+\nu$ is the dominant decay mode. The partial decay width is calculated as \cite{Davidson:2010sf}
\begin{equation}
	\Gamma(H^+\to \ell^+ \nu)=\frac{M_{H^+} \langle m_\nu^2 \rangle_\ell}{8\pi v_2^2},
\end{equation}
where the neutrino final states are summed over. The reweighted neutrino mass squared is defined by
\begin{equation}
	\langle m_\nu^2 \rangle_\ell=\sum_i m_{\nu_i}^2 |U_{\ell i}|^2.
\end{equation}

\begin{figure}
	\begin{center}
		\includegraphics[width=0.49\linewidth]{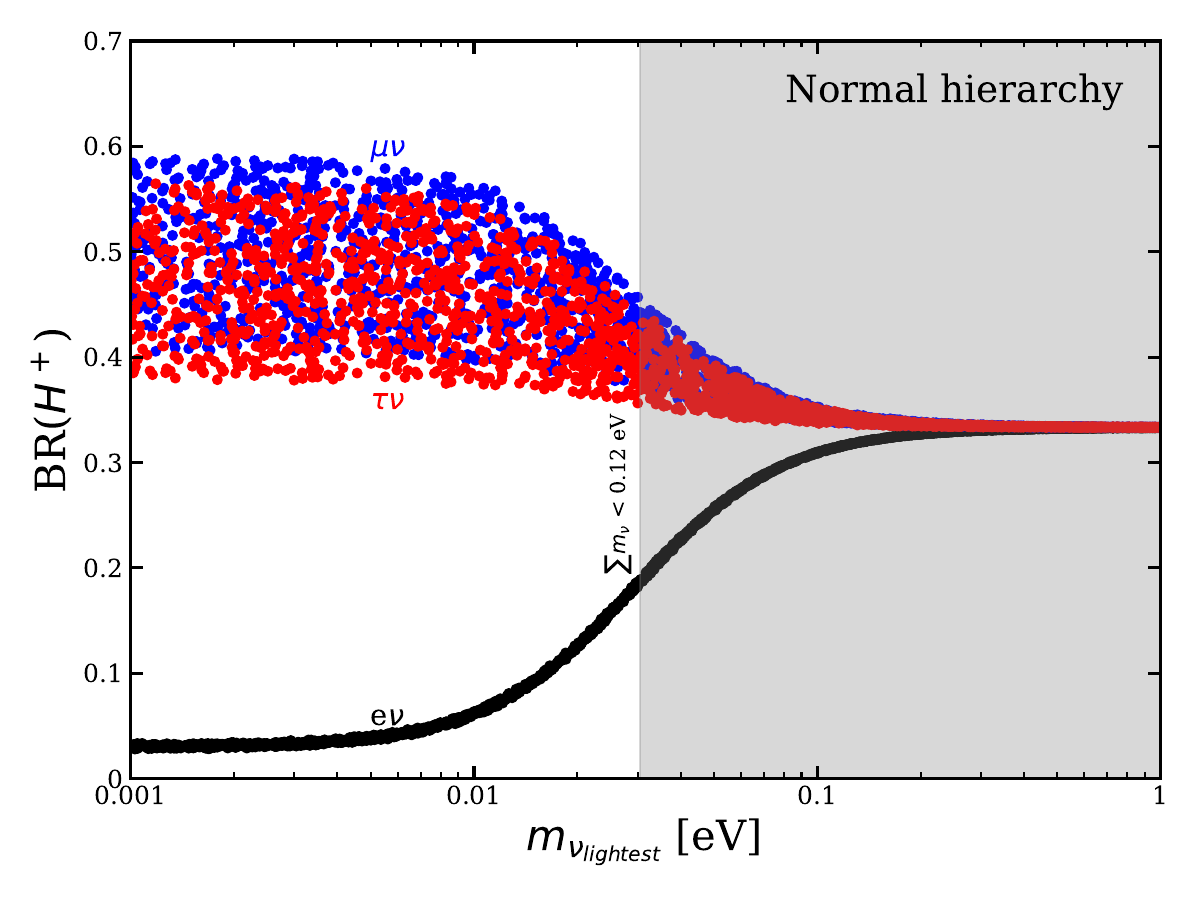}
		\includegraphics[width=0.49\linewidth]{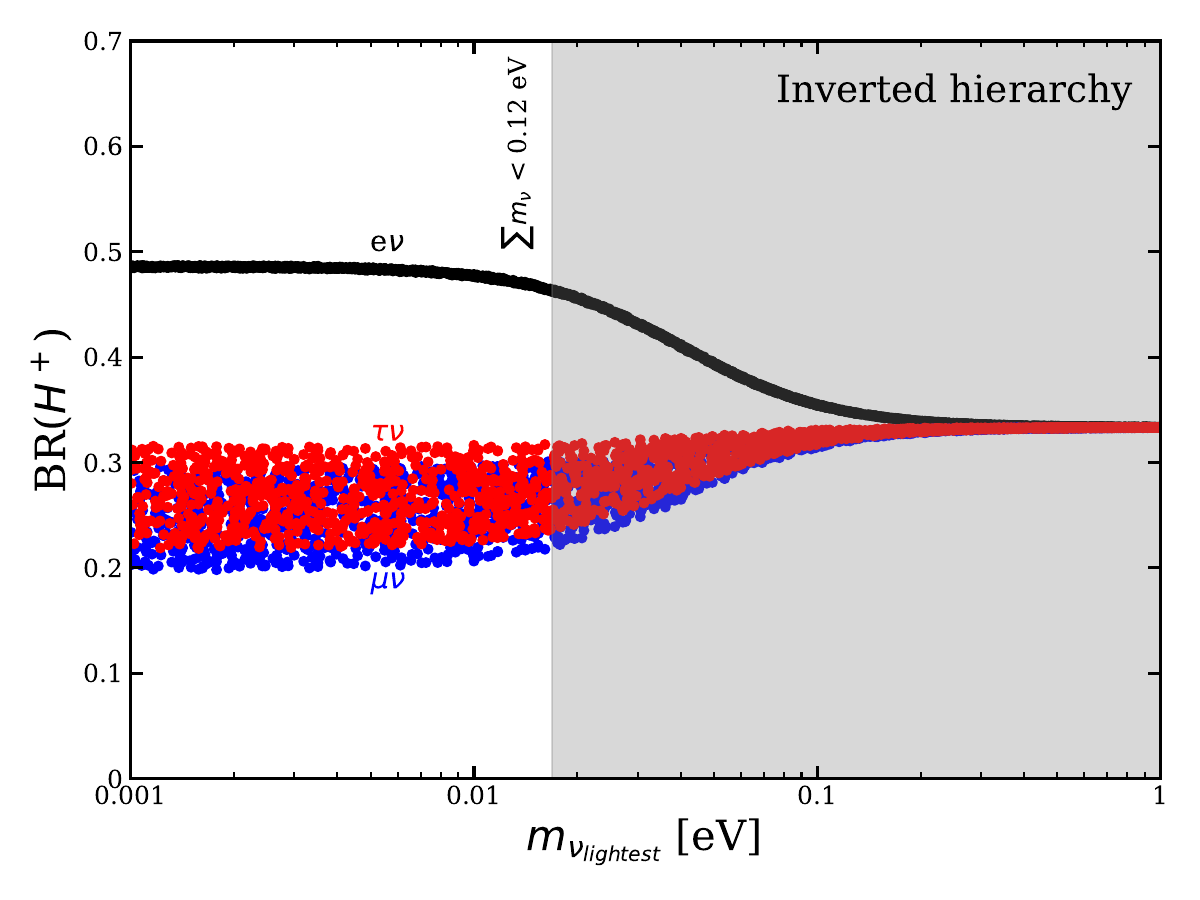}
	\end{center}
	\caption{ Branching ratio of charged scalar $H^+$ as a function of lightest neutrino mass. The shaded region is excluded by cosmological limit $\sum m_\nu < 0.12$ eV.}
	\label{FIG:BRM}
\end{figure}

For a rough estimation, we can use the Tri-Bi-Maximal mixing pattern \cite{Harrison:2002er}, which leads to
\begin{eqnarray}
	\langle m_\nu^2 \rangle_e &\simeq&\frac{2}{3}m_{\nu_1}^2+\frac{1}{3}m_{\nu_2}^2\\
	\langle m_\nu^2 \rangle_\mu =\langle m_\nu^2 \rangle_\tau &\simeq&\frac{1}{6}m_{\nu_1}^2+\frac{1}{3}m_{\nu_2}^2+\frac{1}{2} m_{\nu_3}^2
\end{eqnarray}
Since the absolute neutrino mass and hierarchy are not determined, the variable $\langle m_\nu^2 \rangle_\ell$ also varies. If we further assume the lightest neutrino is massless, then we can derive that
\begin{eqnarray}
	\langle m_\nu^2 \rangle_e^\text{NH}\simeq \frac{1}{3} \Delta m_\text{SOL}^2&,&\langle m_\nu^2 \rangle_\mu^\text{NH} =\langle m_\nu^2 \rangle_\tau^\text{NH}\simeq \frac{1}{2}\Delta m_\text{ATM}^2 ~,\\ 
	\langle m_\nu^2 \rangle_e^\text{IH}\simeq \Delta m_\text{ATM}^2~&,&\langle m_\nu^2 \rangle_\mu^\text{IH} =\langle m_\nu^2 \rangle_\tau^\text{IH}\simeq \frac{1}{2} \Delta m_\text{ATM}^2 ~,
\end{eqnarray} 
where $\Delta m_\text{SOL}^2\simeq7.4\times10^{-5}\text{eV}^2$ and $\Delta m_\text{ATM}^2\simeq2.5\times10^{-3}\text{eV}^2$. These rough estimations predict BR$(H^+\to e^+\nu)\sim \Delta m_\text{SOL}^2/(3\Delta m_\text{ATM}^2)\simeq0.01$, BR$(H^+\to \mu^+\nu)=$BR$(H^+\to \tau^+\nu)\simeq0.5$ for normal hierarchy (NH), and BR$(H^+\to e^+\nu)\simeq 0.5$, BR$(H^+\to \mu^+\nu)=$BR$(H^+\to \tau^+\nu)\simeq0.25$ for inverted hierarchy (IH) respectively. In figure \ref{FIG:BRM}, we show the predicted branching ratio by varying the neutrino oscillation parameters in the $3\sigma$ range. The rough estimations are obviously consistent with the explicit calculation when $m_{\nu_\text{lightest}}\lesssim0.01$ eV.

The neutrino mass hierarchy can be easily distinguished by the measurement of BR$(H^+)$. Because the sum of neutrino masses is constrained as $\sum m_\nu<0.12$ eV, the lightest neutrino mass should be less than 0.03 eV (0.015 eV) for NH (IH). Under this constraint, it is clear in figure \ref{FIG:BRM} that BR$(H^+\to e^+\nu)$ should be less than 0.2 for NH, and decreases as $m_{\nu_\text{lightest}}=m_{\nu_1}$ becomes smaller. When $m_{\nu_\text{lightest}}<0.01$~eV, BR$(H^+\to e^+\nu)\simeq0.03$ is nearly a constant. So if the lightest neutrino mass is in the range of 0.01~eV $<m_{\nu_\text{lightest}}<0.03$ eV, we can probe it by precise measurement of BR$(H^+\to e^+\nu)$. Meanwhile, we have BR$(H^+\to e^+\nu)\simeq0.48$ for IH under the cosmological constraint, and precise measurement of BR$(H^+\to e^+\nu)$ is hard to tell the absolute value of the lightest neutrino mass. 

\begin{figure}
	\begin{center}
		\includegraphics[width=0.49\linewidth]{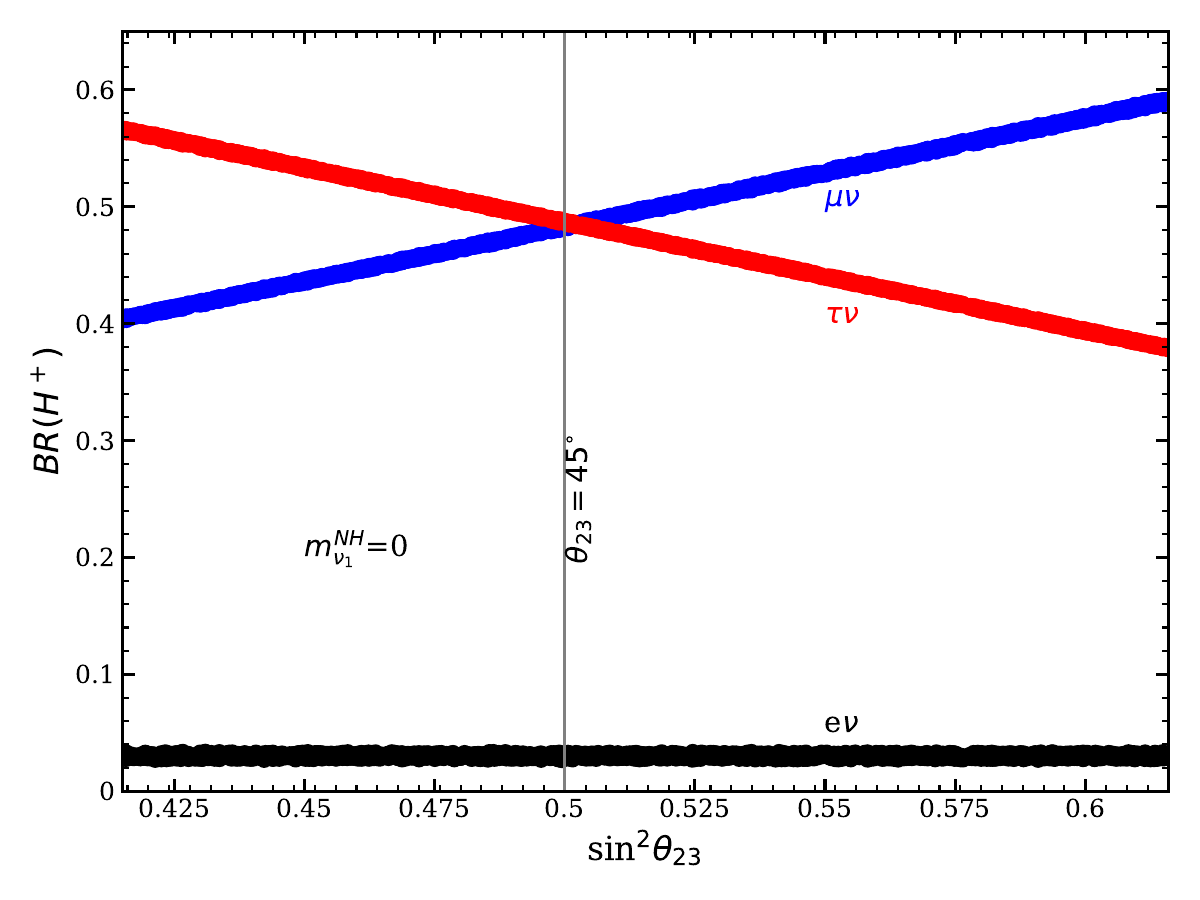}
		\includegraphics[width=0.49\linewidth]{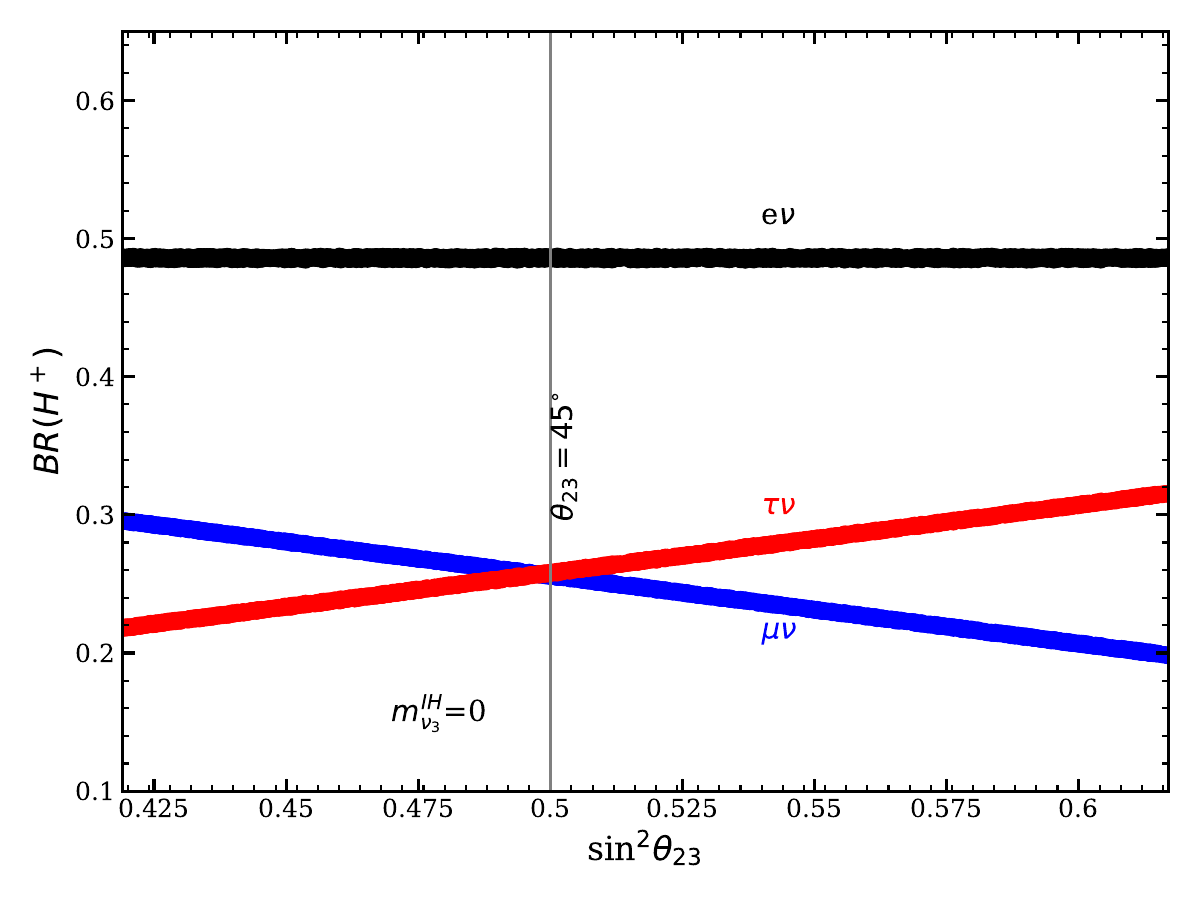}
	\end{center}
	\caption{ Branching ratio of charged scalar $H^+$ as a function of $\sin^2\theta_{23}$ for the normal hierarchy (left panel) and inverted hierarchy (right panel). The lightest neutrino mass is fixed to zero. }
	\label{FIG:BRTh}
\end{figure}

The scanned result shows that BR($H^+\to\mu^+\nu,\tau^+\nu$) is sensitive to the oscillation parameters, while BR$(H^+\to e^+\nu)$ is not. Considering $m_{\nu_\text{lightest}}=0$ and the fact that $\sin^2\theta_{13}\ll1$, the factor $\langle m_\nu^2 \rangle_\ell$ can be approximately calculated as
\begin{eqnarray}
	\langle m_\nu^2 \rangle_e^\text{NH}\simeq \Delta m_\text{SOL}^2 s_{12}^2+\Delta m_\text{ATM}^2s_{13}^2 &,& \langle m_\nu^2 \rangle_\mu^\text{NH}\simeq \Delta m_\text{ATM}^2s^2_{23},\langle m_\nu^2 \rangle_\tau^\text{NH}\simeq \Delta m_\text{ATM}^2c^2_{23}, \label{mvNH}\\
	\langle m_\nu^2 \rangle_e^\text{IH}\simeq \Delta m_\text{ATM}^2 &,&\langle m_\nu^2 \rangle_\mu^\text{IH}\simeq \Delta m_\text{ATM}^2c^2_{23},\langle m_\nu^2 \rangle_\tau^\text{IH}\simeq \Delta m_\text{ATM}^2s^2_{23}. \label{mvIH}
\end{eqnarray}
In the approximation of $\langle m_\nu^2 \rangle_e^\text{NH}$, we have kept the $\Delta m_\text{ATM}^2s_{13}^2$ term because $\Delta m_\text{SOL}^2/\Delta m_\text{ATM}^2\sim s_{13}^2$. From equation \eqref{mvNH} and \eqref{mvIH}, we can derive that $\sum \langle m_\nu^2 \rangle_\ell^\text{NH}\approx \Delta m_\text{ATM}^2$ and $\sum \langle m_\nu^2 \rangle_\ell^\text{IH}\approx 2 \Delta m_\text{ATM}^2$. So for both mass hierarchies, BR$(H^+\to e^+\nu)$ is not sensitive to the oscillation parameters. The branching ratios of the charged scalar as a function of $\sin^2\theta_{23}$ are shown in figure \ref{FIG:BRTh}. According to equation \eqref{mvNH}, BR($H^+\to\mu^+\nu$) increases while BR($H^+\to\tau^+\nu$) decreases as $\sin^2\theta_{23}$ increases for NH. The opposite is the case for IH. Therefore, by precise measurement of BR($H^+\to\mu^+\nu,\tau^+\nu$), the precision of atmospheric mixing angle $\theta_{23}$ can also be improved. 

\section{Dilepton Signature at Colliders}\label{SEC:CLD}

\begin{figure}
	\begin{center}
		\includegraphics[width=0.6\linewidth]{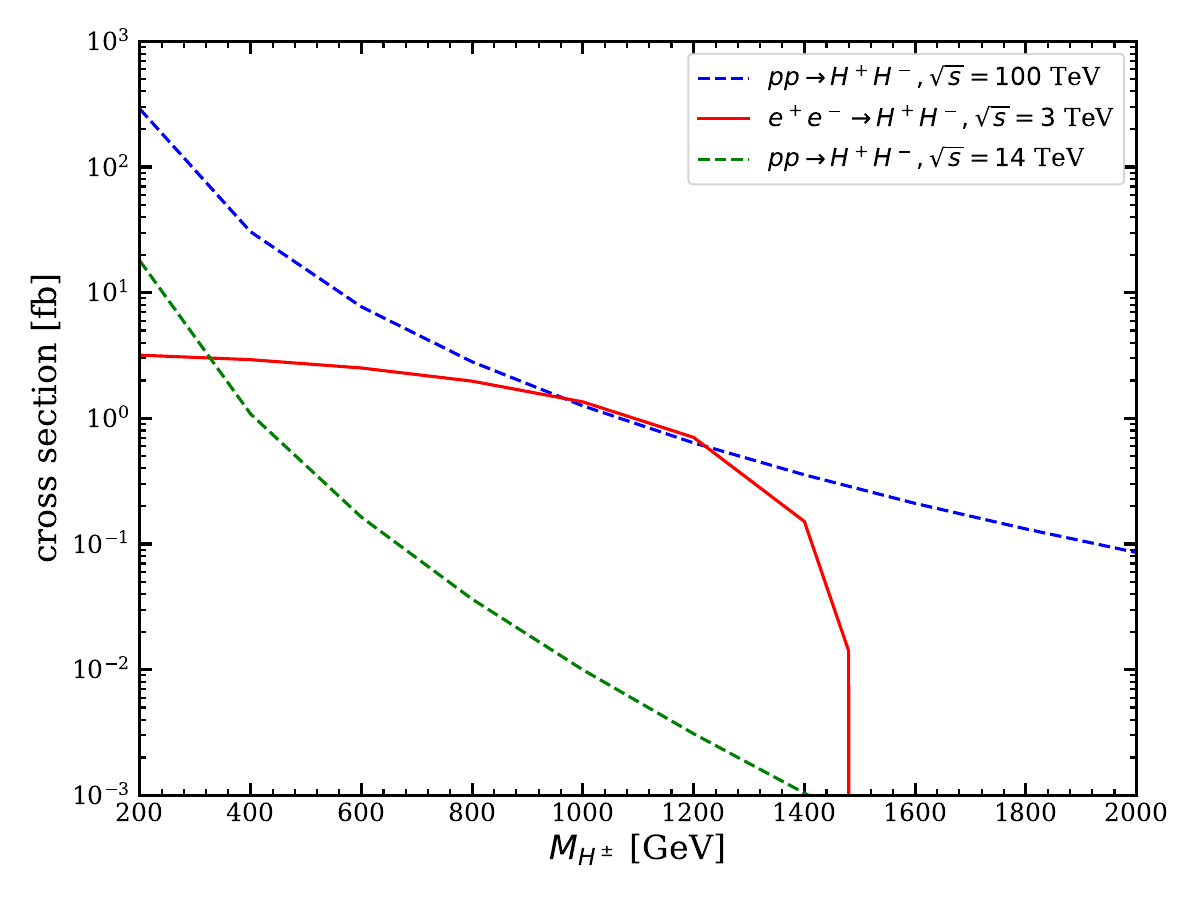}
	\end{center}
	\caption{ Cross section of $H^+H^-$ at the 14 TeV LHC (green dashed) the 3 TeV CLIC (red solid) and the 100 TeV FCC-hh collider (blue dashed).}
	\label{FIG:CS}
\end{figure}

In this section, we study the striking dilepton signature from pair production of charged scalar at colliders. Currently, searches at LHC have excluded $M_{H^+}\lesssim 700$~GeV \cite{ATLAS:2019lff,CMS:2020bfa}. This limit is obtained by assuming  BR$(H^+\to\ell^+\nu)=1$ with $\ell=e,\mu$. In the $\nu$2HDM, we have BR$(H^+\to\ell^+\nu)\approx0.5$ for NH and BR$(H^+\to\ell^+\nu)\approx0.75$ for IH. By a simple cut based analysis, we find that $M_{H^+}\lesssim403$~GeV for NH and $M_{H^+}\lesssim464$~GeV  for IH can be excluded by current experimental searches at 95\% CL. With an integrated luminosity of 3000 fb$^{-1}$,  the projected future LHC result might exclude $M_{H^+}\lesssim760$~GeV for NH and $M_{H^+}\lesssim870$~GeV for IH at 95\% CL.

In this paper, we choose $M_{H^+}=800$ GeV as the benchmark point for the following collider study.  As shown in figure \ref{FIG:CS}, the production cross sections of $H^+H^-$ at the 3 TeV CLIC and 100 TeV FCC-hh are much larger than it at the 14 TeV LHC for TeV scale $H^\pm$.  So we consider the dilepton signature at the former two colliders. 

To ascertain the promising region, we employ MadGraph5\_aMC@NLO \cite{Alwall:2014hca} to generate the background and signal events, which are all at the leading order. Fragmentation and hadronization are incorporated by Pythia8 \cite{Sjostrand:2014zea}. The Delphes3 package \cite{deFavereau:2013fsa} is used to simulate the detector response with the corresponding card for CLIC and FCC-hh. Due to lower tagging efficiency, we do not consider the $\tau$ final state in the following collider simulations.

\subsection{The 3 TeV CLIC}\label{SEC:CLIC}

\begin{figure}
	\begin{center}
		\includegraphics[width=0.45\linewidth]{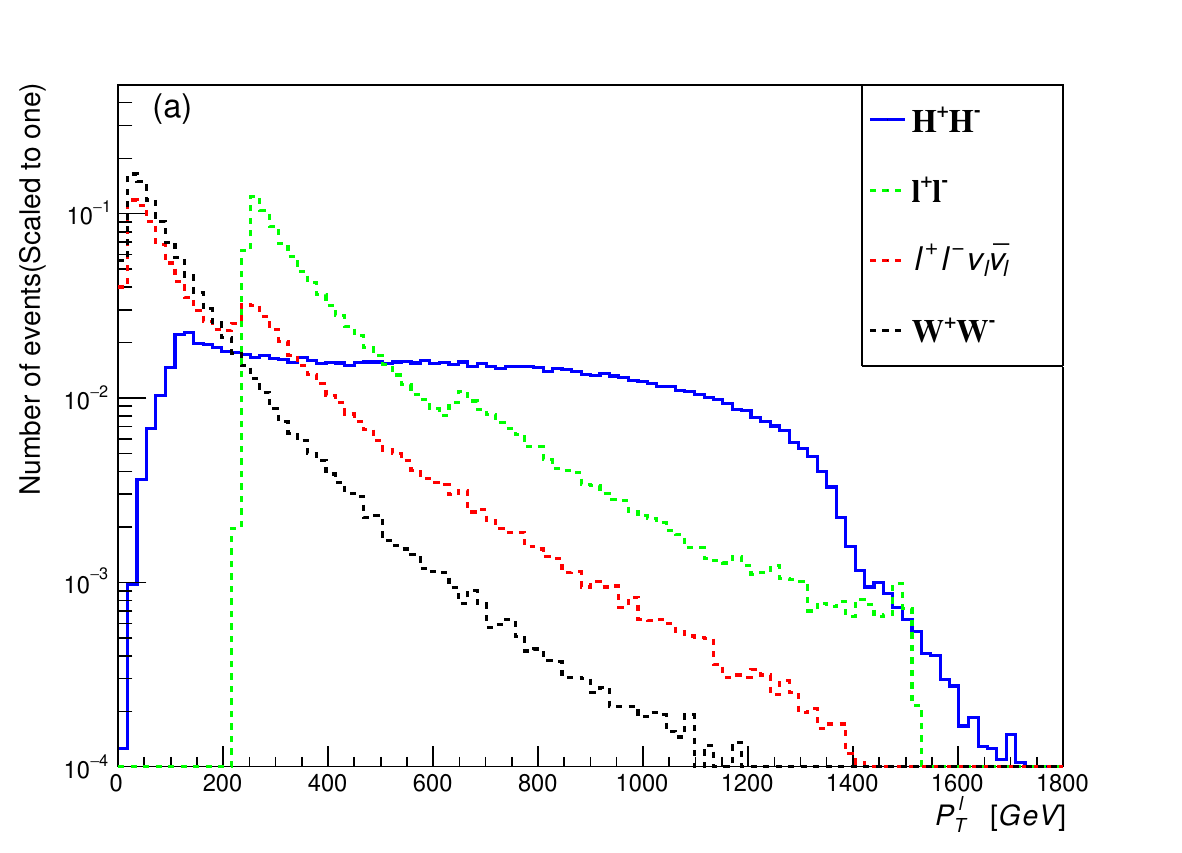}
		\includegraphics[width=0.45\linewidth]{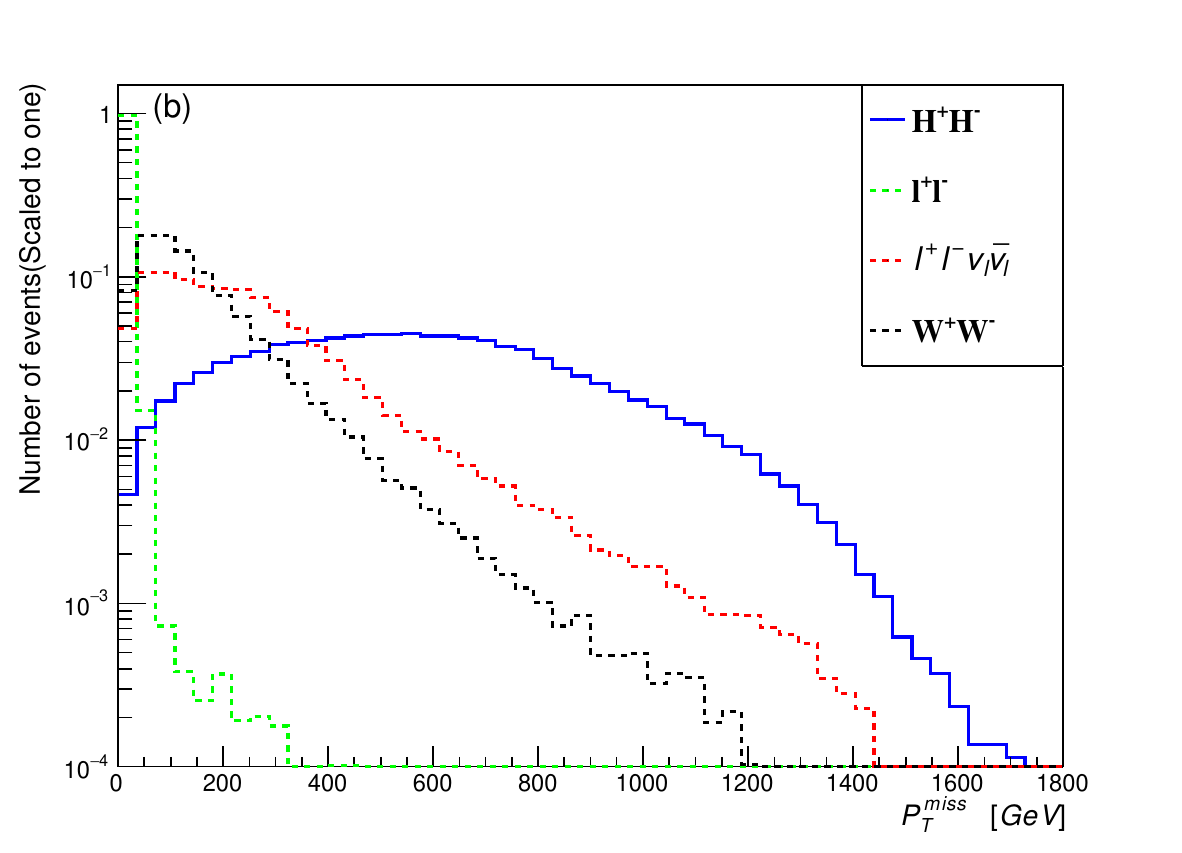}	
	    \includegraphics[width=0.45\linewidth]{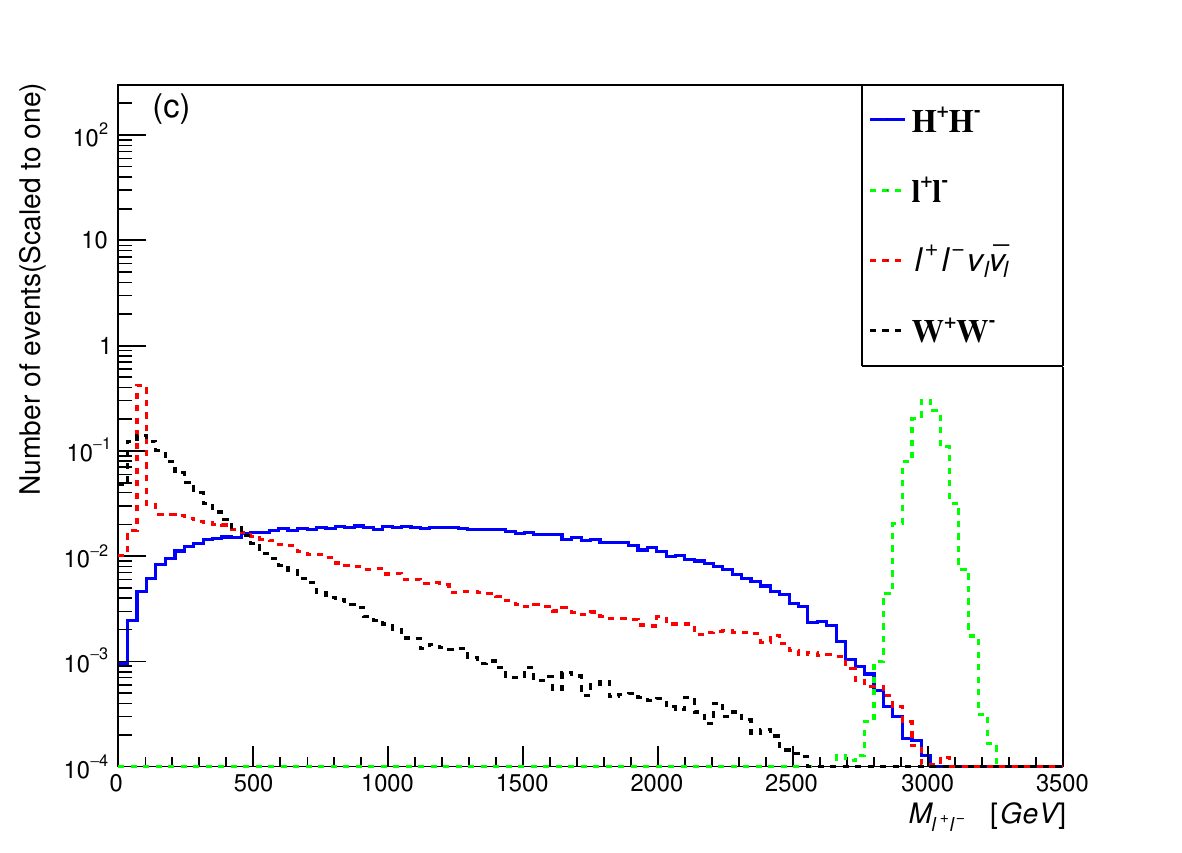}
		\includegraphics[width=0.45\linewidth]{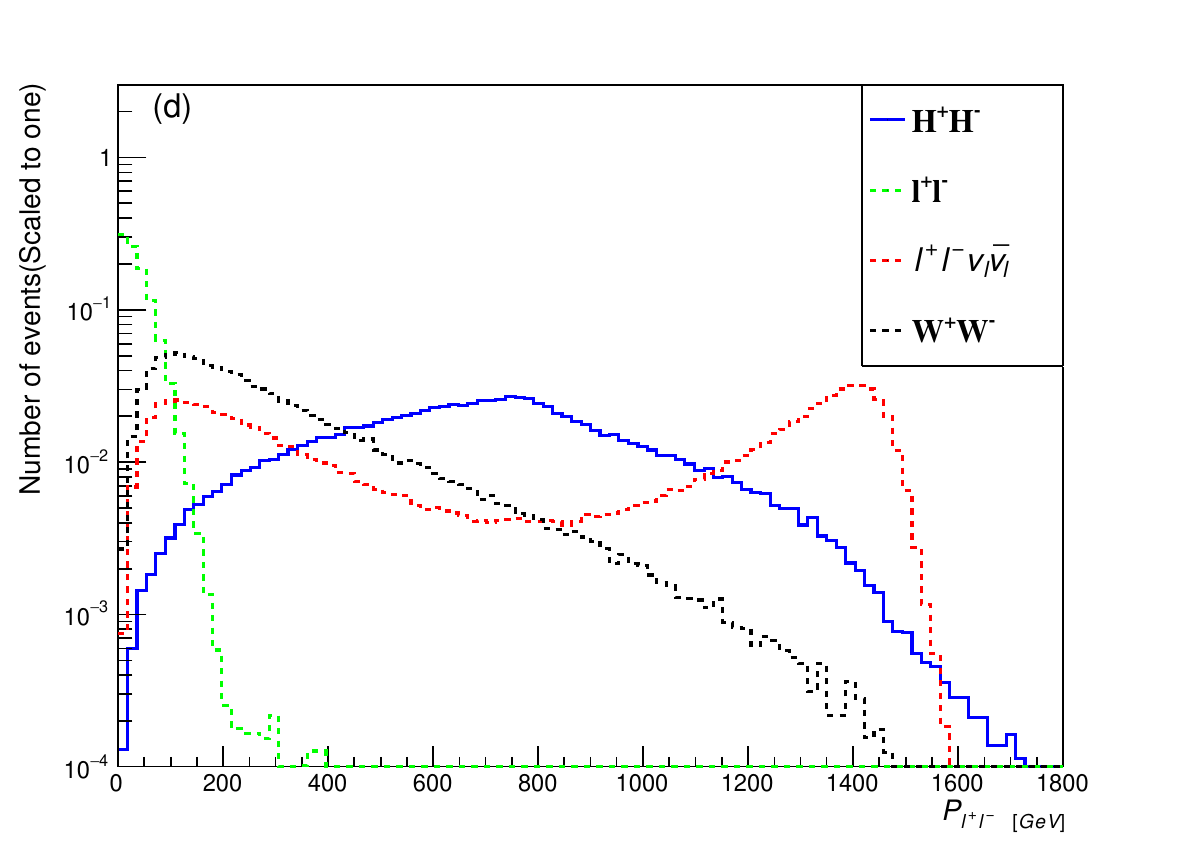}
		\includegraphics[width=0.45\linewidth]{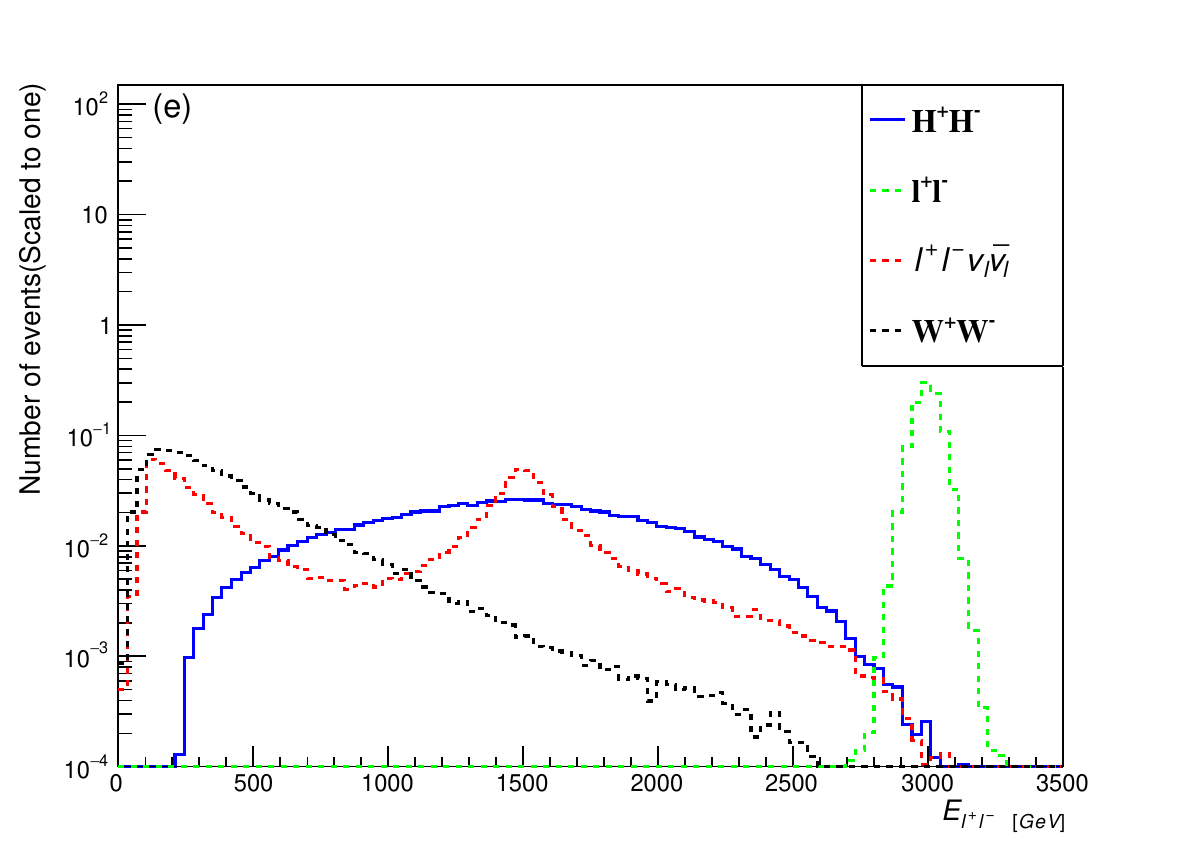}
		\includegraphics[width=0.45\linewidth]{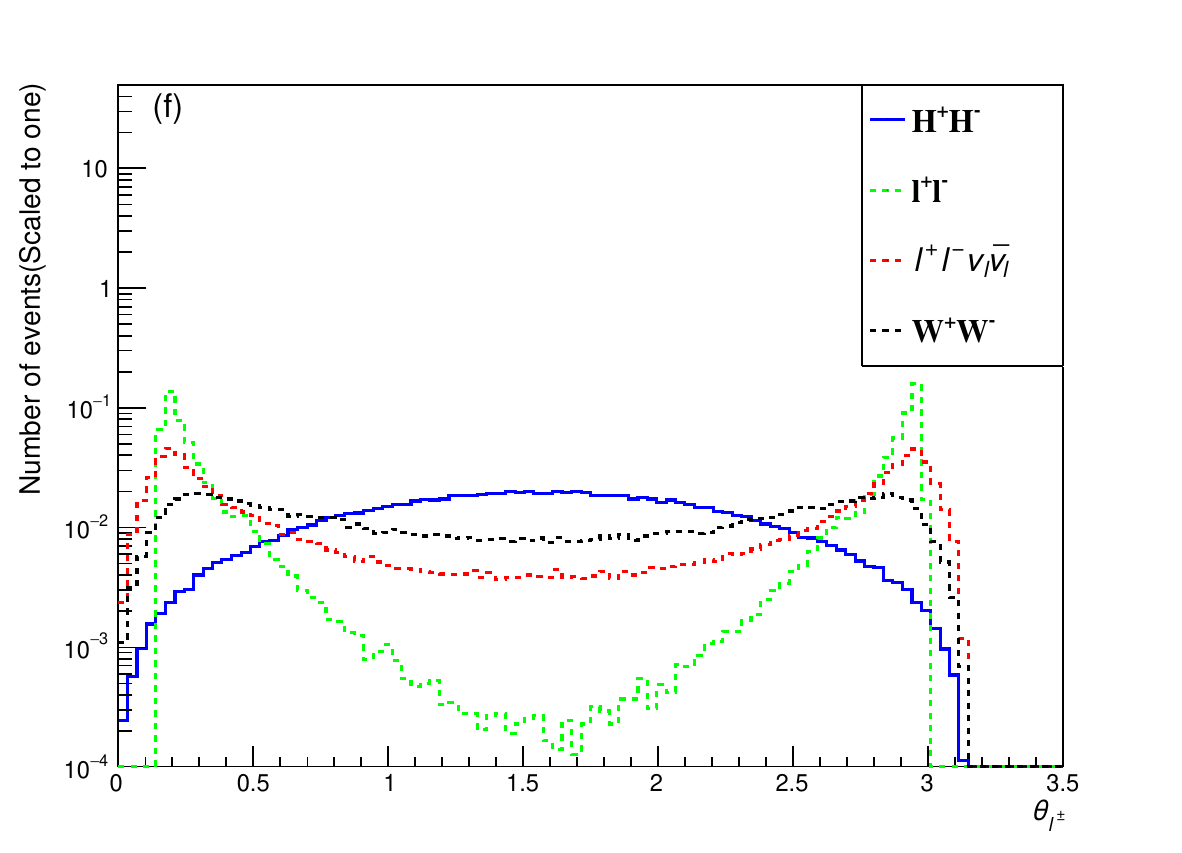}
		\includegraphics[width=0.45\linewidth]{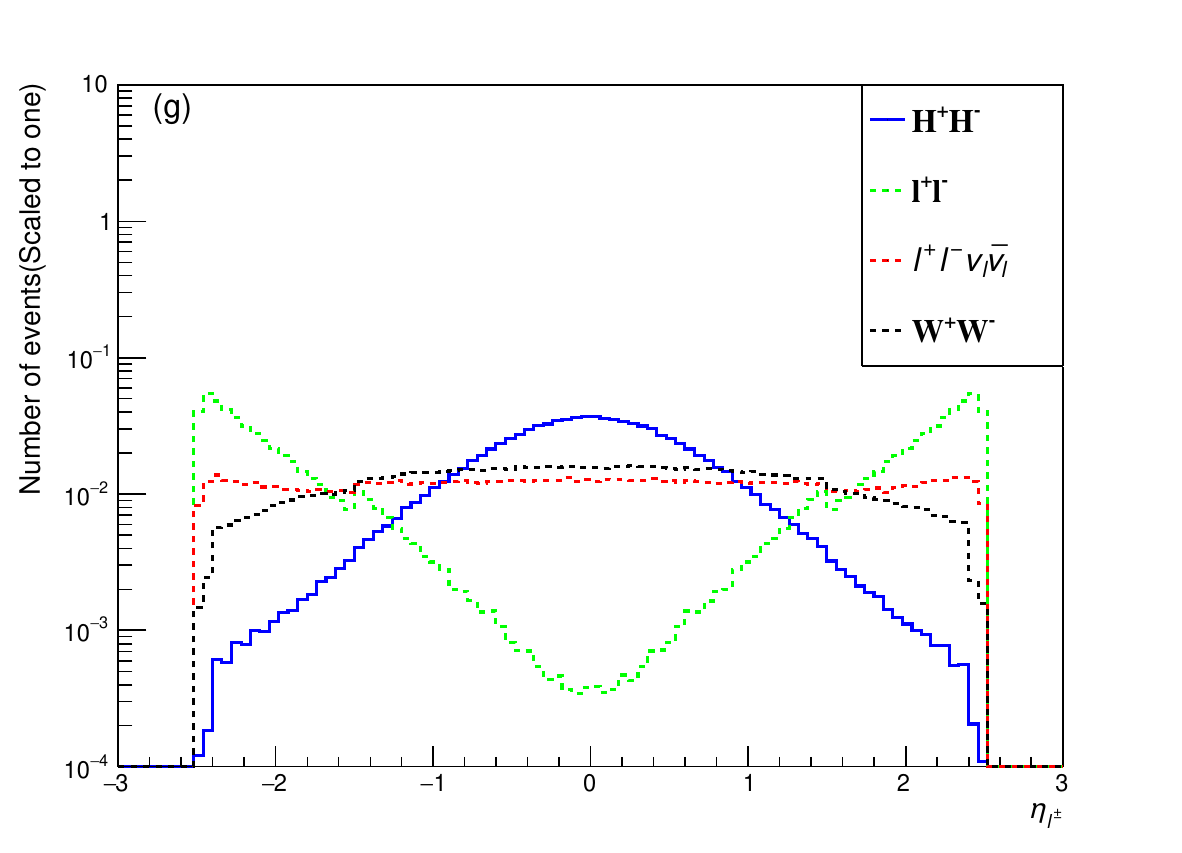}
		\includegraphics[width=0.45\linewidth]{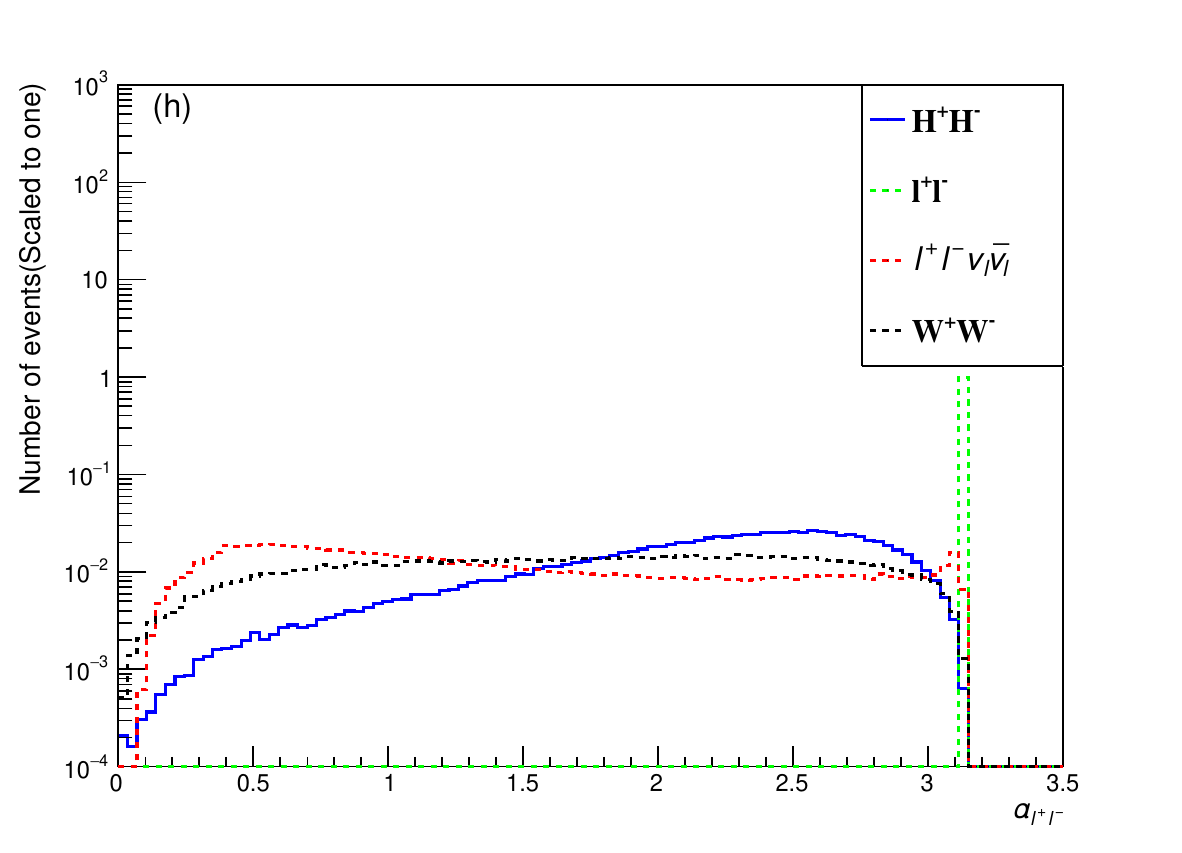}
		\vspace{-2em}
	\end{center}
	\caption{ Normalized distribution of transverse momentum $P_{T}^{\ell}$ (a), missing transverse momentum $P_{T}^{miss}$ (b), dilepton invariant mass $M_{\ell^{+}\ell^{-}}$ (c), vector sum of dilepton momentum $P_{\ell^{+}\ell^{-}}$ (d), dilepton energy $E_{\ell^{+}\ell^{-}}$ (e), polar angle $\theta_{\ell^{\pm}}$ (f), pseudorapidity $\eta_{\ell^{\pm}}$ (g), and dilepton azimuthal angle $\alpha_{\ell^{+}\ell^{-}}$ (h) at the 3 TeV CLIC. }
	\label{FIG:DS1}
\end{figure}

The dilepton signature at the future electron-positron collider CLIC arises from the pair production of charged scalar as
\begin{equation}
	e^+e^-\to H^+ H^- \to \ell^+\nu_\ell + \ell^- \bar{\nu}_\ell\to \ell^+\ell^-+P_T^{miss},
\end{equation}
where $\ell=e,\mu$ and the missing transverse momentum $P_T^{miss}$ corresponds to the contribution of two light neutrinos. The dominant SM backgrounds are
\begin{equation}
	e^+e^-\to\ell^+\ell^-,\ell^+\ell^-\nu_\ell \bar{\nu}_\ell,W^+W^-\nu_\ell \bar{\nu}_\ell
\end{equation} 
Normalized distributions of some variables are shown in figure~\ref{FIG:DS1}. 

In this paper, we perform a simple cut based analysis.  Firstly, we select events with two leptons of opposite charge
\begin{equation}
	N_{\ell^\pm}=1, P_T^\ell>50~\text{GeV}.
\end{equation}
According to figure \ref{FIG:DS1} (a), the cut on $P_T^\ell$ aims to suppress the $\ell^+\ell^-\nu_\ell \bar{\nu}_\ell$ and $W^+W^-\nu_\ell \bar{\nu}_\ell$ backgrounds. The missing transverse momentum $P_T^{miss}$ is another distinct feature. We observe that the signal tends to have a larger $P_T^{miss}$, so we further require relatively large values as
\begin{equation}
	P_T^{miss}>540~\text{GeV}
\end{equation}
This cut is quite efficient to eliminate the $\ell^+\ell^-$ background. After applying these cuts, the dominant background is $\ell^+\ell^-\nu_\ell \bar{\nu}_\ell$. In the distribution of invariant mass of dilepton $M_{\ell^+\ell^-}$, the $\ell^+\ell^-\nu_\ell \bar{\nu}_\ell$ background has a sharp peak around $M_Z$, which indicates that the dilepton originates from $Z\to\ell^+\ell^-$. One may perform a simple $Z$-veto in the invariant mass window between 80 GeV and 100 GeV to suppress the $\ell^+\ell^-\nu_\ell \bar{\nu}_\ell$ background. However, the $M_{\ell^+\ell^-}$ from $W^+W^-\nu_\ell \bar{\nu}_\ell$ channel has a long tail structure. Based on panel (c) of figure~\ref{FIG:DS1}, we then apply the cut
\begin{equation}
	M_{\ell^+\ell^-}>710~\text{GeV}.
\end{equation}

In the distribution of vector sum of dilepton momentum $P_{\ell^+\ell^-}\equiv|\vec{P}_{\ell^+}+\vec{P}_{\ell^-}|$ as in panel (d) of figure~\ref{FIG:DS1}, some events of the $\ell^+\ell^-\nu_\ell \bar{\nu}_\ell$ background could lead to $P_{\ell^+\ell^-}$ larger than the signal. To exclude such background samples, we adopt the cut
\begin{equation}
	P_{\ell^+\ell^-}<980~\text{GeV},
\end{equation}
which is very efficient to eliminate the $\ell^+\ell^-\nu_\ell \bar{\nu}_\ell$ background.

In panel (e) of figure~\ref{FIG:DS1}, it is clear that the sum of dilepton energy $E_{\ell^+ \ell^-}=E_{\ell^+}+E_{\ell^-}$ from the $\ell^+\ell^-$ background is around 3000 GeV. Then the cut
\begin{equation}
	E_{\ell^+\ell^-}<1900~\text{GeV}
\end{equation}
is able to suppress the $\ell^+\ell^-$ background to a negligibly small level. Meanwhile, the $\ell^+\ell^-\nu_\ell \bar{\nu}_\ell$ and $W^+W^-\nu_\ell \bar{\nu}_\ell$ backgrounds need more cuts to suppress. In this paper, we consider the dilepton azimuthal angle $\alpha_{\ell^+\ell^-}$, the pseudorapidity $\eta_{\ell^\pm}$, and lepton polar angle $\theta_{\ell^\pm}$. According to the distributions in panel (f)-(h) of figure~\ref{FIG:DS1},  we then apply the cuts
\begin{equation}
1.5<\alpha_{\ell^+\ell^-}<2.8,~ |\eta_{\ell^\pm}|<1.3,~ 1<\theta_{\ell^\pm}<2.3.
\end{equation}
These cuts are able to make the signal and backgrounds at the same order.

\begin{table}
	\begin{center}
		\begin{tabular}{c| c | c | c |c | c} 
			\hline
			\hline
			$\sigma$(fb) &  $H^+H^-$ (NH) &   $H^+H^-$(IH) &\quad $\ell^+\ell^-$\quad~ & $\ell^+ \ell^-\bar{\nu}_\ell\nu_\ell$ & $W^+W^-\bar{\nu}_\ell \nu_\ell$ \\
			\hline
			$N_{\ell^\pm}=1,P_T^\ell>50$ GeV &  0.322  & 0.553    & 1987 & 104.4  & 1.416 \\ 
			\hline
			$P_T^{miss}>540$ GeV &  0.170  &  0.292   & 0.177 &  10.67 & 0.059 \\
			\hline
			$M_{\ell^+\ell^-}>710$ GeV &  0.126 &   0.216  & 0.177 &  4.983 & 0.021 \\
			\hline
			$P_{\ell^+\ell^-}<980$ GeV &  0.089   & 0.153   & 0.152 &  0.871 & 0.014 \\
			\hline
			$E_{\ell^+\ell^-}<1900$ GeV &  0.067  &  0.114   & 0 &  0.298 & 0.012 \\
			\hline
			$1.5<\alpha_{\ell^+\ell^-}<2.8$   & 0.066   &  0.113   & 0 &  0.216 & 0.009 \\
			\hline
			$|\eta_{\ell^\pm}|<1.3$  & 0.057   &  0.097   & 0 &  0.090 & 0.005 \\
			\hline
			$1<\theta_{\ell^\pm}<2.3$ &  0.048  &  0.083 & 0 &  0.047 & 0.004 \\
			\hline \hline
		\end{tabular}
	\end{center}
	\caption{Cut flow table for the dilepton signal at the 3 TeV CLIC and corresponding backgrounds.
		\label{Tab:CLIC}}
\end{table} 

Results are summarized in table \ref{Tab:CLIC}. During this calculation, we have fixed the neutrino oscillation parameters to the best fit values with zero lightest neutrino mass. With an integrated luminosity of 5000 fb$^{-1}$, the significance can reach $10.8\sigma$ for NH and $16.0\sigma$ for IH respectively. Since BR$(H^\pm\to e^\pm\nu)\ll$BR$(H^\pm\to \mu^\pm\nu)$ for  NH, the dilepton final state is actually dominant by the $\mu^+\mu^-$ pair, and the $e^+e^-,e^\pm\mu^\mp$ pairs are hard to probe. Meanwhile, because of BR$(H^\pm\to e^\pm\nu)\simeq2\times$BR$(H^\pm\to \mu^\pm\nu)$ for IH, the dilepton final state ratio is $e^+e^-:e^\pm\mu^\mp:\mu^+\mu^-\approx4:4:1$. Therefore, the neutrino mass hierarchy can also be determined by the flavor structure of dilepton signature. Such a specific dilepton ratio is useful to distinguish this model from the other neutrino models with dilepton signature \cite{Guella:2016dwo}.

\begin{figure}
	\begin{center}
		\includegraphics[width=0.49\linewidth]{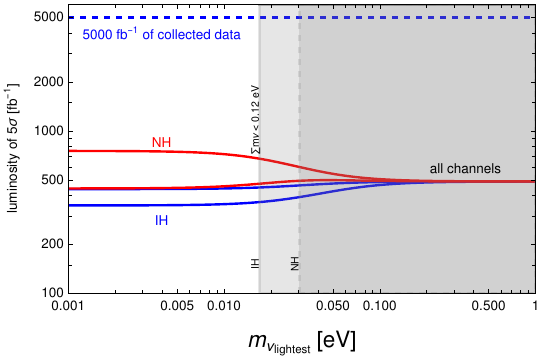}
		\includegraphics[width=0.49\linewidth]{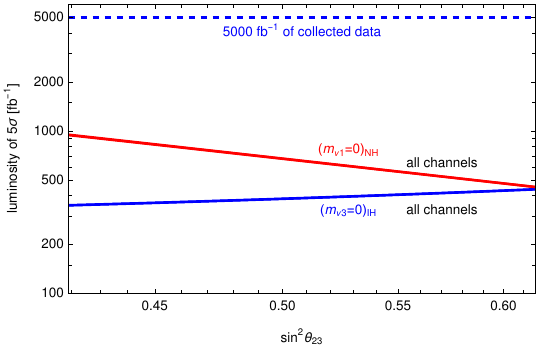}
	\end{center}
	\caption{ Luminosity required at the 3 TeV CLIC for $5\sigma$ discovery. Here $M_{H^\pm}=800$ GeV. }
	\label{FIG:LM}
\end{figure}

Since the branching ratio is affected by neutrino oscillation parameters, we show the required luminosity for $5\sigma$ discovery in figure \ref{FIG:LM}. From the discussion in Section \ref{SEC:BR}, we are aware that the lightest neutrino mass and the mixing angle $\theta_{23}$ have the dominant influence. The lower and upper lines are derived with maximum and minimum predicted values of BR$(H^\pm\to e^\pm\nu,\mu^\pm\nu)$ for both NH and IH respectively. The minimum (maximum) required luminosity for $5\sigma$ discovery is about $450~\text{fb}^{-1}$($760~\text{fb}^{-1}$) for NH and $350~\text{fb}^{-1}$($450~\text{fb}^{-1}$) for IH, respectively. With relatively larger branching ratio into $e$ and $\mu$ final states, we find that the required luminosity of $5\sigma$ discovery for IH is always smaller than that for NH under the constraint of the lightest neutrino mass. The biggest uncertainty comes from the mixing angle $\theta_{23}$, which is clearly shown in right panel of figure \ref{FIG:LM}. The luminosity of $5\sigma$ discovery decreases as $\sin^2\theta_{23}$ increases for NH, while it increases for IH. Using these linear relations, the value of $\sin^2\theta_{23}$ can be improved once there is clear excess in the dilepton signature.

\begin{figure}
	\begin{center}
		\includegraphics[width=0.49\linewidth]{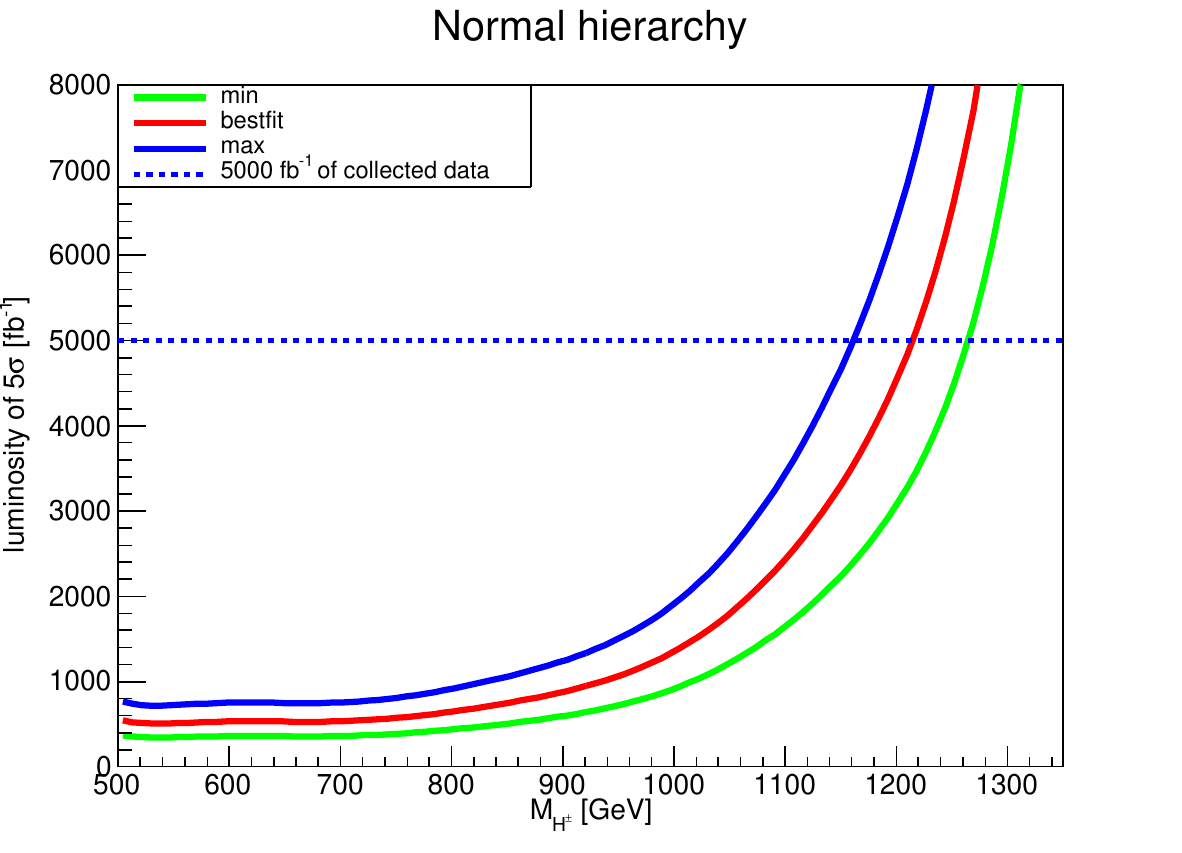}
		\includegraphics[width=0.49\linewidth]{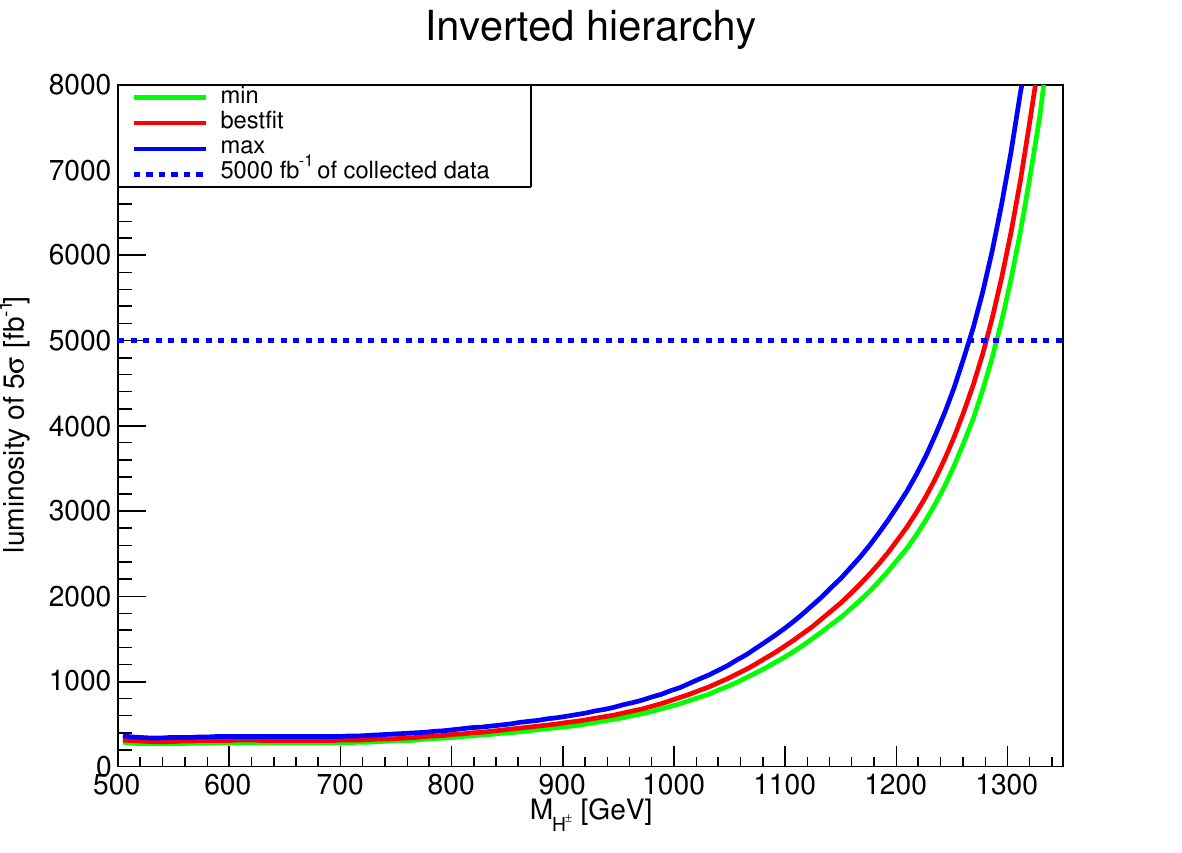}
	\end{center}
	\caption{ Luminosity required at the 3 TeV CLIC for $5\sigma$ discovery as a function of $M_{H^+}$ for  the normal	hierarchy (left panel) and inverted hierarchy (right panel).}
	\label{FIG:LM3}
\end{figure}

In figure \ref{FIG:LM3}, we show the required luminosity  for $5\sigma$ discovery as a function of $M_{H^+}$. Here, we consider $M_{H^+}>500$ GeV to satisfy current LHC dilepton searches with realistic branching ratio. The red lines are derived with best fit neutrino oscillation parameters, while the blue and green lines correspond to the maximum and minimum required values. Compared to IH, the NH is less promising for certain $M_{H^+}$.  With 5000 fb$^{-1}$ integrated luminosity, the 3 TeV CLIC could discover $M_{H^+}\lesssim1220$ GeV for NH, and $M_{H^+}\lesssim1280$ GeV for IH. Based on the result in figure \ref{FIG:CS}, it is obvious that when $M_{H^+}$ is larger than 1300 GeV, the production cross section of $H^+H^-$ at 3 TeV CLIC will be suppressed by the phase space. For heavier $M_{H^+}$, it is expected more promising at the 100 TeV FCC-hh collider.

\subsection{The 100 TeV FCC-hh}\label{SEC:hh}

\begin{figure}
	\begin{center}
		\includegraphics[width=0.49\linewidth]{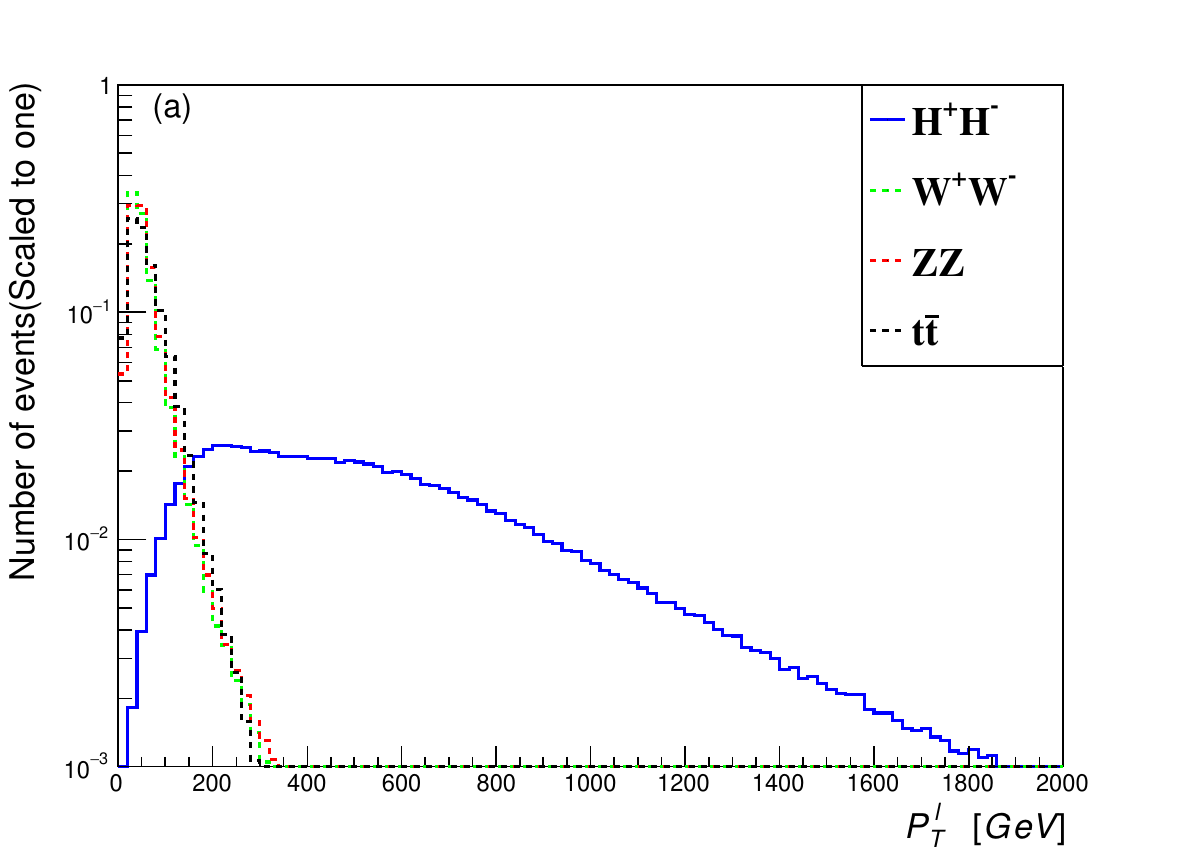}
		\includegraphics[width=0.49\linewidth]{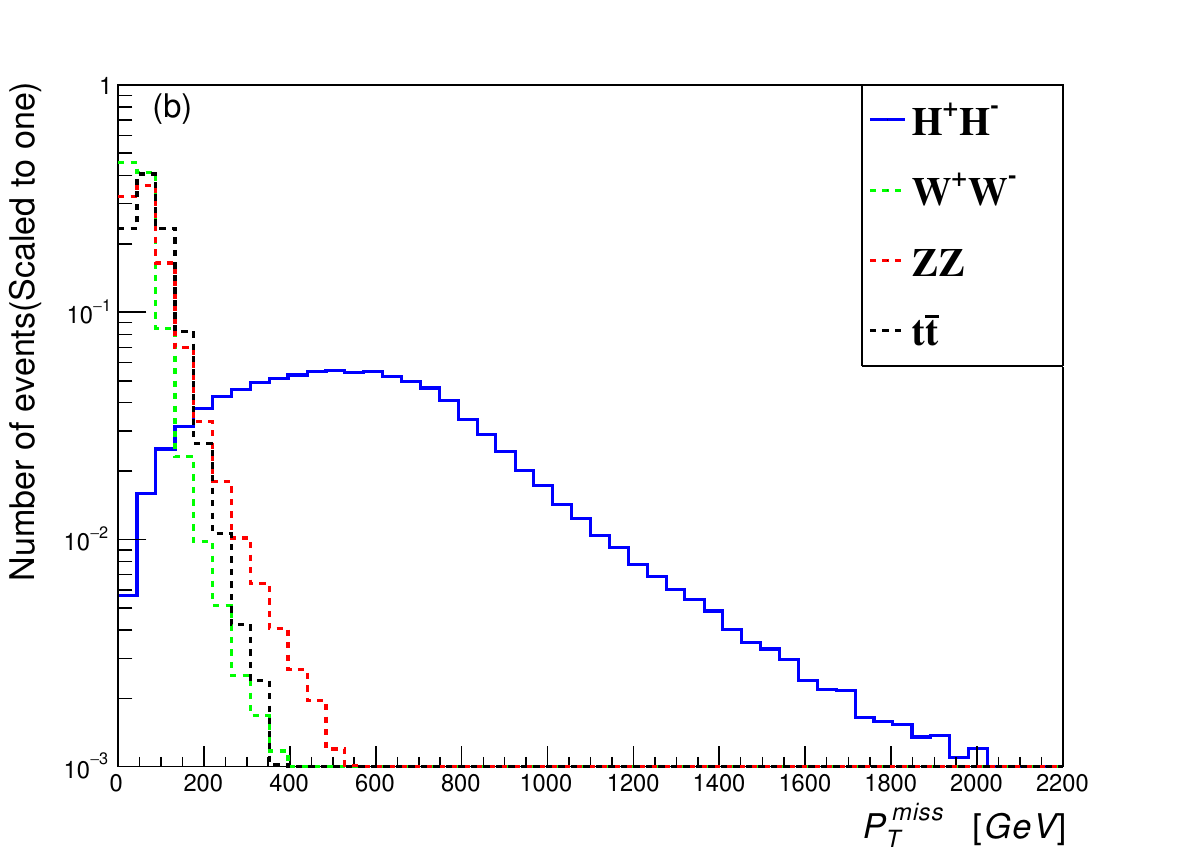}
		\includegraphics[width=0.49\linewidth]{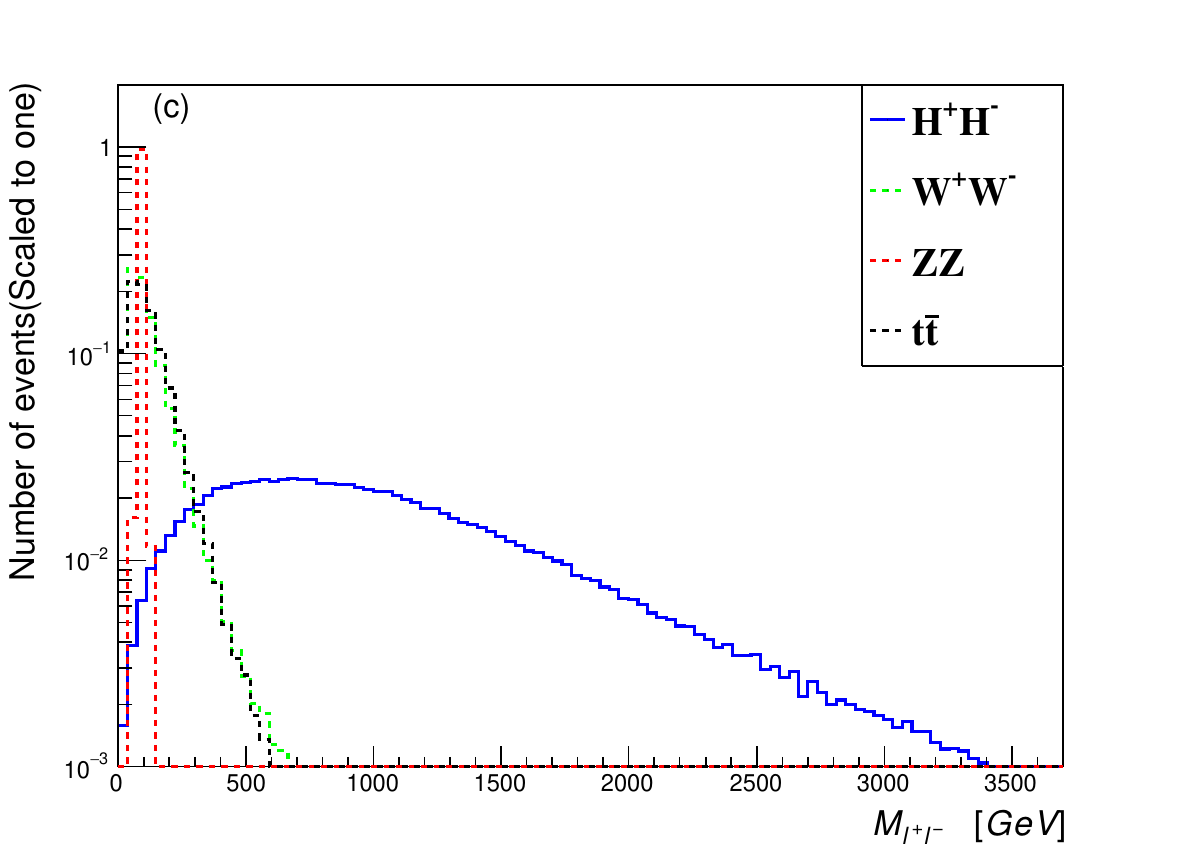}
		\includegraphics[width=0.49\linewidth]{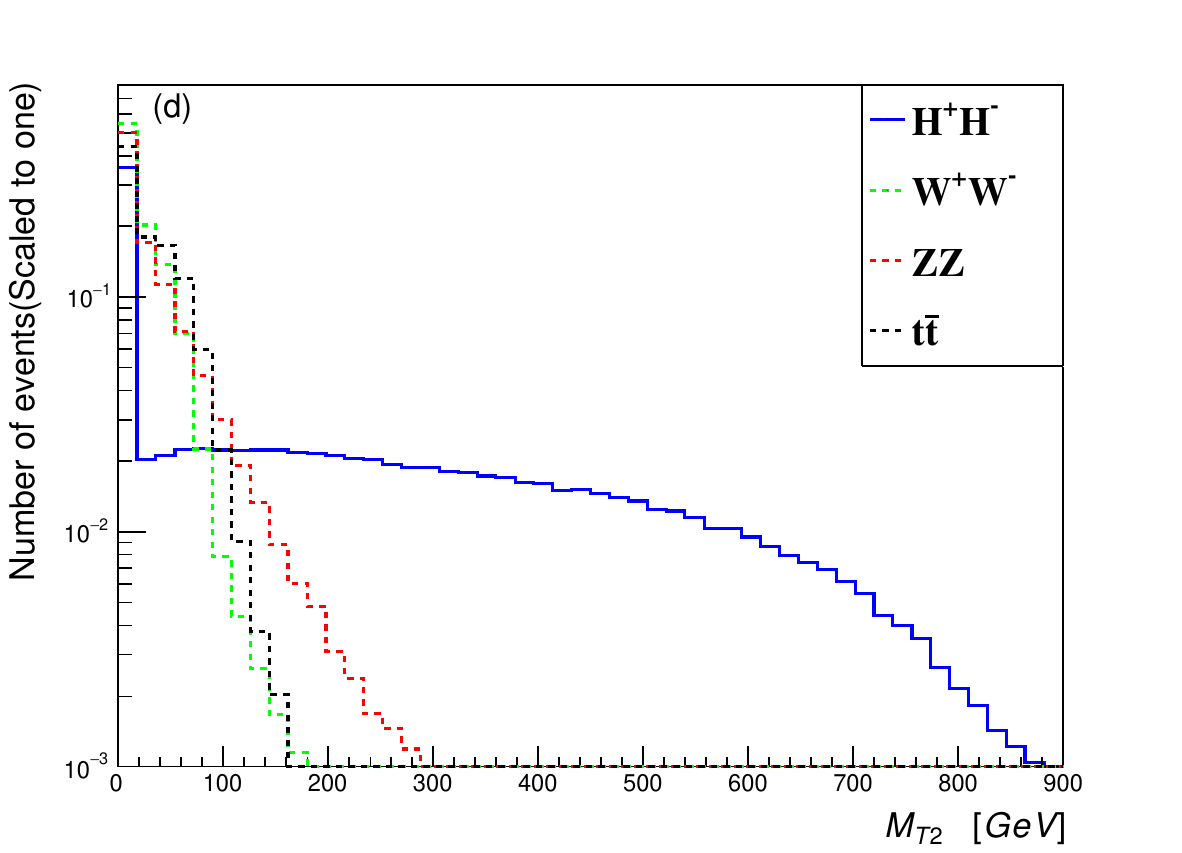}
	\end{center}
	\caption{ Normalized distribution of transverse momentum $P_{T}^{\ell}$ (a), missing transverse momentum $P_{T}^{miss}$ (b), dilepton invariant mass $M_{\ell^{+}\ell^{-}}$ (c) and variable $M_{T2}$ (d) at 100 TeV FCC-hh.}
	\label{FIG:DS2}
\end{figure}

In this section, we investigate the dilepton signature at the planned 100 TeV FCC-hh collider. The signal process is
\begin{equation}
	pp\to H^+H^-\to \ell^+\nu_\ell + \ell^- \bar{\nu}_\ell\to \ell^+\ell^-+P_T^{miss}.
\end{equation}
The dominant SM backgrounds come from
\begin{equation}
	pp\to t\bar{t},W^+W^-,ZZ.
\end{equation}
Normalized distributions of some variables at the 100 TeV FCC-hh collider are shown in figure~\ref{FIG:DS2}.

Similar to the analysis in previous section \ref{SEC:CLIC} for CLIC, we first select events with two opposite sign charged leptons
\begin{equation}
	N_{\ell^\pm}=1.
\end{equation}
The final states of $H^+H^-$ are purely leptonic, whereas the final states of $t\bar{t}$ usually contain several jets from cascade decay $t\to b W$. In order to reduce the enormous $t\bar{t}$ background, events are required to have no energetic jets, i.e.,
\begin{equation}
	N_j =0.
\end{equation}

The leptons from direct decay of charged scalars are typically more energetic than the backgrounds, so a cut on the transverse momentum of leptons is also applied to suppress the SM backgrounds
\begin{equation}
	P_T^{\ell}>100~\text{GeV}
\end{equation}

To obtain events with clear missing transverse momentum, we then require
\begin{equation}
	P_T^{miss}>100~\text{GeV}.
\end{equation}
After implementing these simple selection cuts, the SM backgrounds exhibit a cross section of approximately a few fb at the 100 TeV FCC-hh collider, whereas the signal's cross section is around $\mathcal{O}(0.01)$ fb for $M_{H^+}=800$ GeV.

For the $ZZ$ background, the opposite sign dilepton is mostly from direct $Z\to\ell^+\ell^-$ decay. Then the $ZZ$ background can be suppressed by requiring the invariant mass of dilepton
\begin{equation}
	M_{\ell^+\ell^-}> 130~\text{GeV},
\end{equation}
which is an efficient way to eliminate the $ZZ$ background to a negligible level. Besides the different mass of the decay particle, the pure leptonic decay of $W^+W^-$ channel has the same topology as the signal. Because there are two neutrinos in the final states, we can not directly reconstruct the mass of the charged scalar or $W$ boson. Instead, the $M_{T2}$ variable is one good option to probe the difference between $M_{H^+}$ and $M_{W}$, which is defined as \cite{Lester:1999tx,Barr:2003rg},
\begin{equation}
	M_{T2}  = \underset{\textbf{q}_{T,1} + \textbf{q}_{T,2} =~ \textbf{P}_T^{miss}}{\text{min}} \left\{\text{max}\left[M_{T} (\textbf{P}_{T}^{l_{1}},\textbf{q}_{T,1}),M_{T}(\textbf{P}_{T}^{l_{2}},\textbf{q}_{T,2})\right]\right\},
\end{equation}
where $\textbf{P}_{T}^{l_{1}}$ and $\textbf{P}_{T}^{l_{2}}$ are the transverse momentum vectors of the two leptons, $\textbf{q}_{T,1}$ and $\textbf{q}_{T,2}$ are all possible combinations of two transverse momentum vectors that satisfy $\textbf{q}_{T,1} + \textbf{q}_{T,2} =\textbf{P}_T^{miss}$.  The $M_{T2}$ variable theoretically predicts $M_{T2}<M_{H^+}$ for the signal and $M_{T2}<M_{W}$ for the $W^+W^-$ background. Distributions of $M_{T2}$ are shown in panel (d) of figure \ref{FIG:DS2}. We then apply the cut
\begin{equation}
	M_{T2}>110~\text{GeV}.
\end{equation}

\begin{table}
	\begin{center}
		\begin{tabular}{c| c | c | c |c | c} 
			\hline
			\hline
			$\sigma$(fb) & $H^+H^-$(NH) & $H^+H^-$(IH)  & $t\bar{t}$ & $W^+W^-$ & $ZZ$ \\
			\hline
			$N_{\ell^\pm}=1$ &  0.4773  &  0.8193   &~ 226442~  & ~ 4789.35 ~ & ~543.54~  \\ 
			\hline
			$N_{j}=0$ & 0.0364   & 0.0626  & 337.74  & 1342.47   &  143.74 \\ 
			\hline
			$P_T^\ell>100$ GeV & 0.0348   &  0.0597   & 19.998  &  44.029  &  1.0857 \\ 
			\hline
			$P_T^{miss}>100$ GeV  & 0.0336  &  0.0577   &  2.222 & 2.4192   & 1.0857  \\
			\hline
			$M_{\ell^+\ell^-}>130$ GeV & 0.0331   &  0.0568   & 2.222  &  2.4192  & 0  \\
			\hline
			$M_{T2}>110$ GeV  &  0.0178  &  0.0305   & 0  & 0   &  0 \\
			\hline \hline
		\end{tabular}
	\end{center}
	\caption{Cut flow table for the dilepton signal at 100 TeV FCC-hh and corresponding backgrounds.
		\label{Tab:FCC}}
\end{table} 

\begin{figure}
	\begin{center}
		\includegraphics[width=0.48\linewidth]{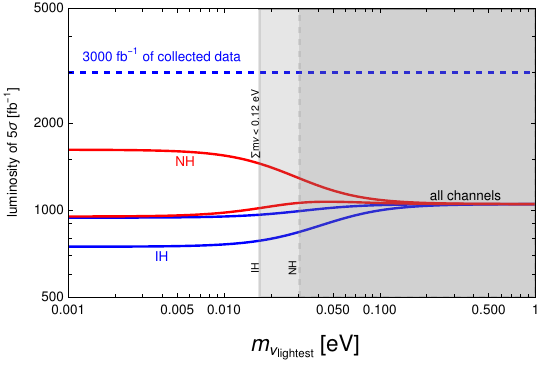}
		\includegraphics[width=0.48\linewidth]{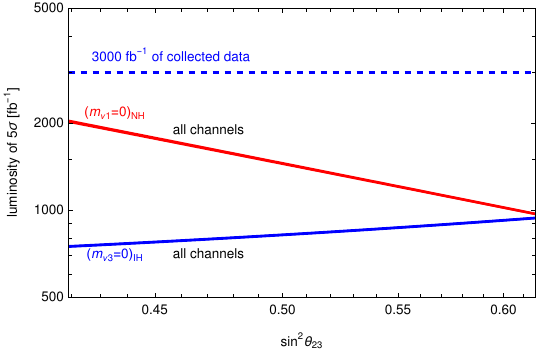}
	\end{center}
	\caption{ Luminosity required at the 100 TeV FCC-hh for $5\sigma$ discovery. Here $M_{H^\pm}=800$ GeV.}
	\label{FIG:FC}
\end{figure}

In table \ref{Tab:FCC}, we summarize the efficiency for each cut, where the cross sections of signal are obtained by fixing the neutrino oscillation parameters to the best fit values with zero lightest neutrino mass. These cuts are able to suppress the SM backgrounds to zero, while keeping the cross section of signal at the order of 0.01~fb. With an integrated luminosity of 3000 fb$^{-1}$, the significance can reach $7.3\sigma$ for NH and $9.6\sigma$ for IH respectively. The required luminosity for $5\sigma$ discovery is about $1400~\text{fb}^{-1}$ for NH and $820~\text{fb}^{-1}$ for IH. Therefore, the future 100 TeV FCC-hh collider needs more integrated luminosity than the 3 TeV CLIC to discover the charged scalar with $M_{H^+}=800$ GeV.

In figure \ref{FIG:FC}, we show the impact of lightest neutrino mass and mixing angle $\theta_{23}$ on the $5\sigma$ discovery luminosity at 100 TeV FCC-hh. For NH, the maximum required luminosity is about $1600~\text{fb}^{-1}$, and the minimum is about $950~\text{fb}^{-1}$. For IH, the required luminosity varies from $750~\text{fb}^{-1}$ to $950~\text{fb}^{-1}$. The dependence of required luminosity on $\sin^2\theta_{23}$ for FCC-hh is the same as the CLIC.

\begin{figure}
	\begin{center}
		\includegraphics[width=0.49\linewidth]{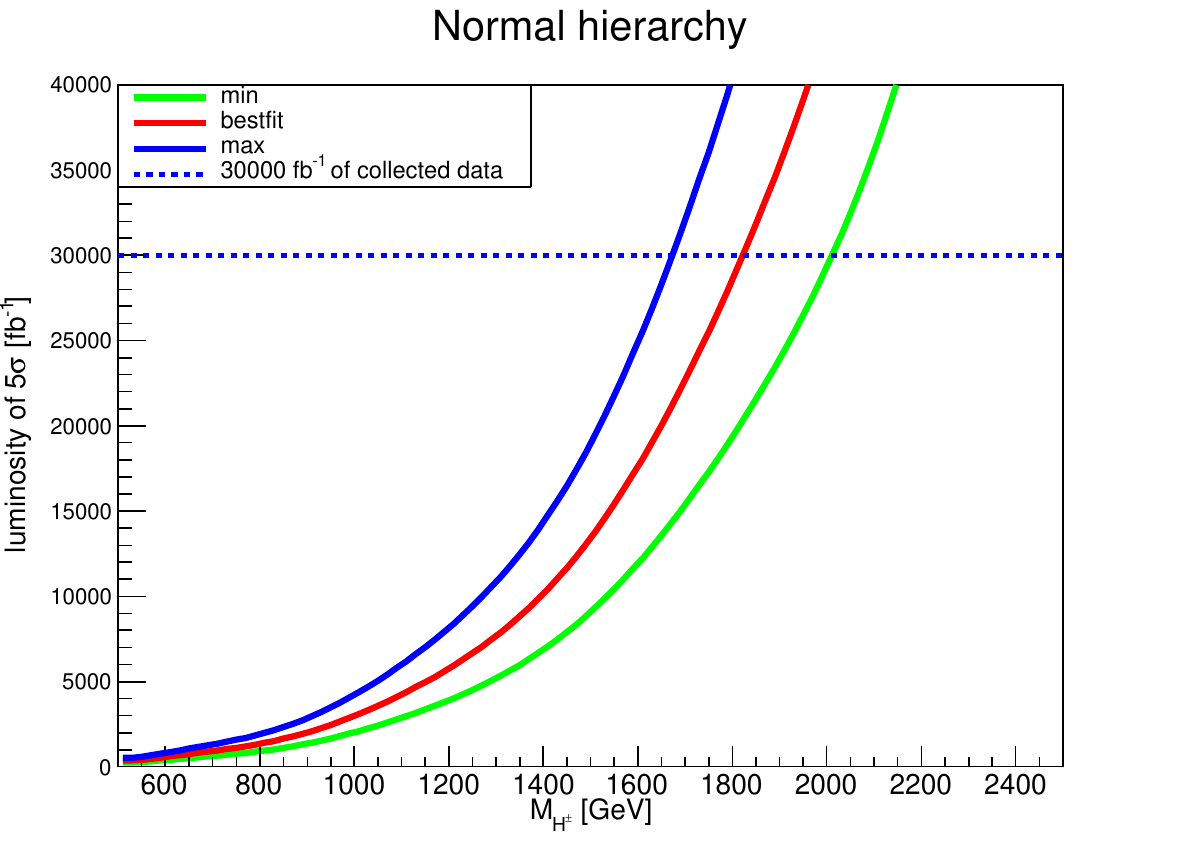}
		\includegraphics[width=0.49\linewidth]{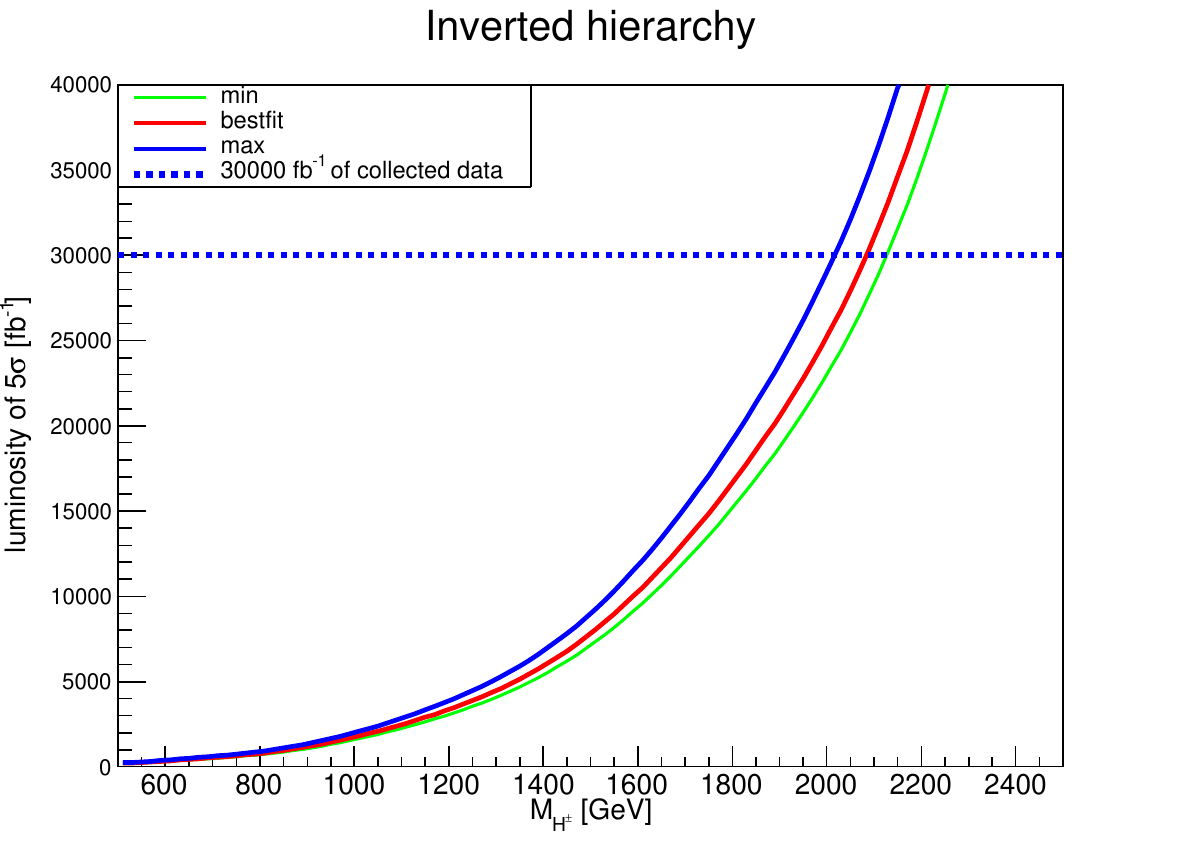}
	\end{center}
	\caption{ Luminosity required at the 100 TeV FCC-hh for $5\sigma$ discovery as a function of $M_{H^+}$ for the normal hierarchy (left panel) and inverted hierarchy (right panel).}
	\label{FIG:FCC3}
\end{figure}

Based on the cuts in table \ref{Tab:FCC}, we then explore the feasibility of dilepton signature as a function of $M_{H^+}$. The results are shown in figure \ref{FIG:FCC3}. With an integrated luminosity of 30 ab$^{-1}$, the future 100 TeV FCC-hh collider could probe $M_{H^+}\lesssim1810$ GeV for NH and $M_{H^+}\lesssim2060$ GeV for IH. Impacted by the uncertainties of neutrino oscillation parameters, the upper discovery value of $M_{H^+}$ for NH would vary from 1600 GeV to 2000 GeV, while for IH, the impact is relatively smaller.

\section{Conclusion}\label{SEC:CL}

The Majorana or Dirac nature of neutrinos is still undetermined. If neutrinos are Dirac particles, the neutrinophilic doublet $\Phi_2$ with eV scale VEV can naturally generate tiny neutrino masses. Such small VEV comes from the soft $U(1)$ broken term $m_{12}^2 \Phi^\dag_1\Phi_2$. Under the $U(1)$ charge assignment $Q_{\Phi_2}=Q_{\nu_R}=1$, this global symmetry forbids both the Yukawa interaction of right-handed neutrinos to SM Higgs and the Majorana mass term of right-handed neutrinos, but allows the interaction between the new Higgs $\Phi_2$ and right-handed neutrinos.

Because the generation of neutrino mass and decay of charged scalar both involve the same Yukawa coupling $y_\nu$, the leptonic branching ratio of charged scalar $H^\pm$ has a strong correlation with the neutrino oscillation parameters. The two parameters that can obviously affect the branching ratio are the lightest neutrino mass and the  atmospheric mixing angle $\theta_{23}$. When the lightest neutrino mass is below 0.01 eV, the NH predicts BR$(H^+\to e^+\nu)\ll$ BR$(H^+\to \mu^+\nu)=$ BR$(H^+\to \tau^+\nu)\simeq0.5$, while the IH predicts BR$(H^+\to e^+\nu)/2\simeq$ BR$(H^+\to \mu^+\nu)\simeq$ BR$(H^+\to \tau^+\nu)\simeq0.25$. Under the cosmological constraint $\sum m_\nu < 0.12$~eV, we can probe the absolute lightest neutrino mass via precise measurement of BR$(H^+\to e^+\nu)$ for NH when $m_{\nu_1}\gtrsim0.01$ eV. However, BR$(H^+\to e^+\nu)$ is nearly a constant for IH. On the other hand, BR($H^+\to\mu^+\nu$) is greatly affected by the mixing angle $\theta_{23}$. Therefore, the precise measurement of BR($H^+\to\mu^+\nu$) can indicate the value of $\theta_{23}$.

The charged scalar can be pair produced at colliders, which then leads to the dilepton signature $\ell^+\ell^-+P_T^{miss}$ with $\ell=e,\mu$. Although current LHC has excluded $M_{H^+}\lesssim700$ GeV with BR($H^+\to \ell^+\nu$)=1, we find the actual limits should be $M_{H^+}\lesssim 403$ GeV for NH and $M_{H^+}\lesssim 464$ GeV for IH with more realistic branching ratios, respectively. In this paper, we then perform a detailed simulation of this signature and corresponding background at the 3~TeV CLIC and the 100 TeV FCC-hh colliders. For fixed value of $M_{H^+}$, the required $5\sigma$ discovery luminosity of the NH scenario is always larger than the IH scenario. The 3 TeV CLIC could reach $5\sigma$ discovery when $M_{H^+}\lesssim1220$~GeV for NH and $M_{H^+}\lesssim1280$ GeV for IH with 3 ab$^{-1}$ data. Meanwhile, the future 100 TeV FCC-hh collider could probe $M_{H^+}\lesssim1810$ GeV for NH and $M_{H^+}\lesssim2060$~GeV for IH with 30 ab$^{-1}$ data.

\section*{Acknowledgments}
This work is supported by the National Natural Science Foundation of China under Grant No. 11605074 and  No. 11805081, Natural Science Foundation of Shandong Province under Grant No. ZR2019QA021 and ZR2022MA056, the Open Project of Guangxi Key Laboratory of Nuclear Physics and Nuclear Technology under Grant No. NLK2021-07, Joint Large-Scale Scientific Facility Funds of the NSFC and CAS under Contracts Nos. U1732263 and U2032115.

\end{document}